\documentclass[a4paper,onecolumn,11pt]{quantumarticle}
\pdfoutput=1
\usepackage[utf8]{inputenc}
\usepackage[english]{babel}
\usepackage[T1]{fontenc}
\usepackage{amsmath}
\usepackage{hyperref}

\usepackage{tikz}
\usepackage{lipsum}

\usepackage{cite}
\usepackage{natbib}
\usepackage[braket, qm]{qcircuit}

\usepackage{graphicx}
\usepackage{float}
\usepackage{caption}
\usepackage{placeins}

\title{A pathway to accurate potential energy curves on NISQ devices}
\author{Ryan Ward}
\affiliation{E.~A.~Milne Centre for Astrophysics, Department of Physics and Mathematics, University of Hull, HU6 7RX, United Kingdom}

\author{David M.~Benoit}
\orcid{0000-0002-7773-6863}
\affiliation{E.~A.~Milne Centre for Astrophysics, Department of Physics and Mathematics, University of Hull, HU6 7RX, United Kingdom}

\author{Francesco Benfenati}
\affiliation{Zapata Computing Inc., 100 Federal Street, Boston, MA 02110, USA}

\date{September 2022}

\begin{document}

\maketitle

\begin{abstract}
    We present a practical workflow to compute the potential energy curve of the hydrogen molecule on near intermediate-scale quantum (NISQ) devices. The proposed approach uses an extrapolation scheme to deliver, with only few qubits, full configuration interaction results close to the basis-set limit. We show that despite the limitations imposed by the noisy nature of simulated quantum hardware, it is possible to recover realistic electronic correlation values, if we also estimate expectation values of the Hartree--Fock ground state energy. Using two models of noisy quantum experiments, we evaluate the performance of a scheme that requires at most a double-zeta basis set (3-21G, in this case) and compare with the most accurate Born--Oppenheimer potential energy curves available in the literature. Our flexible approach is implemented using simple variational ansatzes combined with straightforward mitigation techniques and thus we expect it to be also suitable for other energy estimation quantum schemes.
\end{abstract}

\section{Introduction}

Current quantum computing hardware is able to perform very elaborate quantum operations on an increasingly large number of qubits. Yet, the applications of quantum computing to quantum chemistry calculations remains more a curiosity than a mass phenomenon. This is mainly due to the difficulty in performing operations on a large number of qubits since many complex gates and high circuit depth are required to perform useful chemical simulations that can rival classical computations. 

Hybrid approaches such as the variational quantum eigensolver (VQE) \citep{Peruzzo:2014} have shown impressive results when modelling electronic structure on quantum computers and thus this algorithm has become a de-facto standard \citep{Elfving:2020,Claudino:2020}. However, for large chemically-relevant systems, the wave function description (triple or quadruple zeta basis sets) and qubit requirements remain too high for most Noisy Intermediate-Scale Quantum (NISQ) devices. As a result, there has been many adaptations of the VQE algorithm to improve both circuit depth and noise reduction (see \cite{jouzdani:2020}, \cite{kottmann:2020} or \cite{benfenati:2021}, for example) - nevertheless most demonstrations remain limited to ``toy" systems. Indeed, a true quantum advantage can only really be achieved by producing results that are useful to the quantum chemistry community and potentially more difficult to obtain on a classical computer than a quantum one. 

One particular issue of current VQE demonstrations is the sub-minimal basis set used to describe the molecular wave function. In a recent paper, \citet{Elfving:2020} highlight that a computation on the hydrogen molecule which begins to approach chemical accuracy would require 56 spin-orbitals in its molecular wave function (cc-pVTZ basis set) or 56 qubits. Although the actual number of necessary qubit depends on the specific implementation of the algorithm and can be lowered slightly using parity, symmetry operations and tapering \citep{bravyi2017tapering}. To be chemically meaningful, most other molecules would require much larger wave functions combined with extrapolation schemes. 

In the present study, we embrace the limitations of quantum computing and instead seek to use the relatively low level basis sets to extract information on the electronic correlation that VQE can estimate and use it to reconstruct high-quality results. This approach is already used commonly (albeit with larger basis sets) in quantum chemistry to obtain reliable estimates of the basis-set limit (also called complete basis set limit, CBS, corresponding to an "infinitely flexible" wave function). A recent paper by \cite{Varandas2018} demonstrated how to use this approach for sub-minimal basis sets on classical computers, if appropriate care is taken in computing a suitable extrapolation function. \\

In the present study, we explore two aspects of the quantum computation of electronic structure: \begin{enumerate}
 \item we apply a dual-level basis set approach to high-accuracy electronic structure calculations using two types of NISQ device simulations and use this approach to obtain an accurate dissociation curves for the H$_2$ molecule. Those curves are compared to existing benchmark results by \citet{sims:2006} for H$_2$, 
 obtained on classical computers.
 \item We develop suitable mitigation techniques that include using "internal standards" and an analytical representation of electronic correlation to minimise the impact of noisy measurements on our proposed approach. This proposed workflow aims to provide a practical blueprint for the determination of smooth and high-accuracy potential energy surfaces. 
\end{enumerate}

This paper is organised as follows: in Sec.~\ref{sec:dual}, we present the dual-level basis set approach used throughout this study, followed by a detailed description of the computational details of our quantum simulations in Sec.~\ref{sec:compdet}. We present our first set of results in Sec.~\ref{sec:generic} and compare our results to the latest available literature data. In Sec.~\ref{sec:noisy}, we explore the application of our approach to a more realistic noisy model of a quantum device and derive a suitable workflow for actual NISQ systems. Finally, we present our conclusions in Sec.~\ref{sec:conc}. 

\section{Method}
\label{sec:dual}

Highly accurate electronic structure calculations on molecular systems is often a trade-off between description of the electronic correlation used (i.e.\ ranging from mean-field Hartree--Fock theory to full configuration interaction, FCI, with many useful approximate treatments in between) and the size of the basis set employed to represent the molecular wave function. One convenient way of approaching the problem is to use an incremental approach, both for electronic correlation and the basis set expansion. This has lead to vast progress over decades and has enabled large-scale, yet accurate, calculations for a number of systems \citep{dunning:1989,Bakowies:2007,Boschen:2017}. 
In this study, we focus on the highest level of electronic correlation, i.e.\ FCI as it provides a numerically exact solution to the Schr{\"o}dinger equation. The formal \emph{factorial} computational scaling of FCI in terms of molecular size limits its applications, but can be implemented in a straightforward manner on quantum computers. In this work, we use a variational quantum eigensolver (VQE) approach with a unitary coupled-cluster (UCCSD) anzatz \citep{Peruzzo:2014} and a hardware-optimised ansatz \citep{Kandala2017} to obtain the FCI ground state for the electronic Hamiltonian expressed in second-quantisation formalism (see also Sec.~\ref{sec:compdet} for details). 

The second aspect of the calculation, namely the limited size of the electronic basis set, is usually side-stepped by performing a number of similar calculations using increasingly larger basis sets. Indeed, \cite{dunning:1989} developed hierarchic basis sets, where each each angular momentum component is individually saturated, and organised them in increasing order of cardinality (2 for double-zeta, 3 for triple-zeta, etc). They showed that the correlation energy follows an exponential curve as a function of cardinal number, as this number can be identified with $L_{max}$, the maximum angular momentum quantum number of the basis set \citep{Halkier:1998}. This approach has been used extensively in the literature on classical computers to estimate the complete basis-set limit (or CBS limit) for both correlated and Hartree--Fock calculations (see \cite{dunning:1989} and \cite{Jensen2005}, for example).

Yet, the smallest basis set used in a Dunning-type CBS expansion is typically a double-zeta basis set (often cc-pVDZ, \citep{dunning:1989}), which would already require on the order of 20 qubits on a quantum computer for a molecule such as H$_2$. The next step in this expansion for H$_2$ would be a triple zeta basis set (typically cc-pVTZ, \citep{dunning:1989}) which requires a staggering 56 qubits, as stated in the introduction. This implies that even the smallest Dunning-type extrapolation would require a very large quantum computer, let alone mitigating the error accumulated due to the resulting VQE circuit depth. 

However, over the past few years, there has been a ``revival" of low-cost methods that typically use much smaller basis sets - e.g.\ minimal or sub-minimal basis sets. The predictions of these models are usually dramatically improved with the addition of simple empirical corrections. This effort has been spearheaded by the Grimme group with the advent of the HF-3c approach \citep{Sure:2013} and followed by other similar techniques \citep{grimme:2015, Brandenburg:2018}. Recently, \citet{Varandas2018} created a bridge between the world of large basis set extrapolation and that of low-cost sub-minimal basis sets, by suggesting an extrapolation method that uses effective cardinal numbers instead. This new development enables the use of CBS extrapolation with much smaller basis sets and thus potentially offers a way to harness the FCI capabilities of quantum computing even on today's NISQ devices. This is what we explore in the present study.

\subsection{Basis set extrapolation}
\label{sec:CBSth}
We define the correlation energy $E^\textrm{corr}$ as the difference between the fully-correlated energy (FCI in the present case) and the mean-field energy obtained through a Hartree--Fock (HF) calculation:
\begin{equation}
 E^\textrm{corr}=E^\textrm{FCI}-E^\textrm{HF}
\end{equation}

In this study, we extrapolate the Hartree--Fock energy using an exponential expression and HF results obtained with Jensen's PC-n basis set \citep{Jensen:2001}. This is performed on a classical computer and uses cardinal numbers 2, 3, 4 and 5 (in other words: pc-1, pc-2, pc-3 and pc-4 basis sets). The extrapolation formula used for this is given by \citep{Jensen2005}:
\begin{equation}
 E^\textrm{HF}(L_{max})=E^\textrm{HF}_\infty+B \exp(-C\cdot L_{max})
 \label{eq:hfenergy}
\end{equation}
Where $E^\textrm{HF}(L_{max})$ is the Hartree--Fock energy obtained with a basis set of largest angular momentum $L_{max}$, $E^\textrm{HF}_\infty$ is the complete basis set (CBS) estimation of the Hartree--Fock energy and $B$ and $C$ are fitting coefficient. This extrapolated HF energy enables us to estimate the "exact" correlation energy by subtracting $E^\textrm{HF}_\infty$ from the numerical results of \citet{sims:2006} for H$_2$. The extrapolated HF energy also serves as a zero-approximation for the construction of our energy curves.

The correlation energy is then computed using FCI calculations performed with two minimal basis sets on a simulated NISQ device combined with the extrapolation procedure suggested by \citet{Varandas2018}:
\begin{equation}
 E^\textrm{corr}_\infty=E^\textrm{corr}_{B}+ \frac{x_B^{-3}}{x_A^{-3}-x_B^{-3}} \left(E^\textrm{corr}_{B}-E^\textrm{corr}_{A}\right)
 \label{eq:correnergy}
\end{equation}
where $E^\textrm{corr}_\infty$ is the CBS estimation of the correlation energy. $E^\textrm{corr}_{i}$ is the correlation energy obtained with basis set $i$ and $x_i$ is the effective cardinal number for basis set $i$ (taken from the averaged values of Table~2 in \citet{Varandas2018}). Finally, $A$ and $B$ are the different minimal basis set used, with basis set $B$ being a larger basis set than $A$.

The CBS-extrapolated total energy, $ E_\infty(r)$, along the dissociation curve is then computed using:
\begin{equation}
 E_\infty(r)=E^\textrm{HF}_\infty(r)+E^\textrm{corr}_\infty(r)
 \label{eq:totenergy}
\end{equation}
for each inter-nuclear distance $r$.

\subsection{Quantum calculation methodology}
Following the method outlined above (Sec.~\ref{sec:CBSth}), we use a classical computer to compute the values of $E^\textrm{HF}_\infty(r)$, as this is a computation that is fast and typically of $\mathcal{O}(N^3)$ complexity (depending on implementation, see \citep{Echenique:2007} for a review). We then use a NISQ device to compute the values of $E^\textrm{corr}_\infty(r)$ at FCI level using two small minimal basis sets, thus keeping the number of qubits needed small. Our choice for this study is basis sets from the MINI family \citep{duijneveldt1971a} and the 3-21G basis set \citep{binkley1980a},

which keep qubit count to a minimum (4 qubits for MINI and 8 qubits for 3-21G, but this can be lowered to 2 and 6 after parity-reduction mapping, for example). Those calculations are performed using the VQE approach \citep{Peruzzo:2014}, where we first construct an electronic Hamiltonian in second-quantisation formalism and translate it to qubit operations using a Bravyi--Kitaev transformation \citep{Seeley2012}, before performing a classical optimisation of the variational ansatz. More details are given in Sec.~\ref{sec:compdet} below.

\section{Computational Details}
In order to construct the potential energy (or dissociation) curve for H$_2$, 
we perform the same series of classical/quantum computations at fixed inter-nuclear distances, $r$. The results are then combined using Eqns.~\ref{eq:hfenergy}, \ref{eq:correnergy} and \ref{eq:totenergy} to generate the curves discussed in the results sections.

\label{sec:compdet}
\subsection{HF extrapolation details}
We use the least-square fitting routines implemented in the \textsc{SciPy} python package \citep{2020SciPy-NMeth} to fit Eq.~\ref{eq:hfenergy} to results obtained using the 4 different basis sets discussed in Sec.~\ref{sec:CBSth}, namely pc-1 to pc-4. All HF calculations for this extrapolation are performed using the orca 5.0.1 package \citep{neese_2012} on a classical computer. We also estimate the error on $E^\textrm{HF}_\infty(r)$ as 1 standard deviation (1$\sigma$) of the fitting parameter.

\subsection{Correlation energy calculation}
The \textsc{psi4} package \citep{psi4:2012} is used to compute the one- and two-body integrals ($h_{ij}$ and $h_{ijkl}$, respectively) needed to build a fermionic Hamiltonian within the Born--Oppenheimer approximation and expressed in a second-quantisation formalism as:
\begin{equation}
 \hat{H}_\mathrm{elec}(r)=\sum_{i,j}h_{ij}\hat{a}_i^\dag\hat{a}_j+\sum_{i,j,k,l}h_{ijkl}\hat{a}_i^\dag\hat{a}_j^\dag\hat{a}_k\hat{a}_l
 \label{eq:hamilto}
\end{equation}
where $\hat{a}_i^\dag$ and $\hat{a}_i$ represent the creation and annihilation operators of the molecular spin-orbital $i$. 

Note that the MINI and 3-21G basis sets both use Cartesian primitives orbitals. While the 3-21G basis set is standard, the MINI basis was added manually to \textsc{psi4} using data obtained from the EMSL database \citep{pritchard2019a,feller1996a,schuchardt2007a}. 

The correlation energy estimate for each basis set, $x$, at a given inter-nuclear separation, $r$, is then computed using:
\begin{equation}
 E^\mathrm{corr}_{x}(r)=E^\mathrm{FCI}_{x}(r)-E^\mathrm{HF}_{x}(r)
 \label{eq:correnergylevelx}
\end{equation}
where the FCI energy, $E^\mathrm{FCI}_{x}$, for basis set $x$, is obtained through either a quantum VQE computation or a reference classical calculation. The HF energy is directly extracted from the \textsc{psi4} results during the generation of the second-quantisation Hamiltonian. 

All classical FCI reference values for both MINI and 3-21G energy curves are computed on a classical computer using the full configuration--interaction code implemented in \textsc{psi4}.

\subsection{Quantum computing simulations}
In order to assess the applicability of our approach, we perform a two types of noisy simulations using the \textsc{Orquestra} software from Zapata Computing Inc.~\citep{orquestra}. The operators described in Eq.~\ref{eq:hamilto} are mapped to Pauli operators using the Bravyi-Kitaev transformation \citep{BRAVYI2002210, Seeley2012} and implemented as quantum circuits. For the first set of simulations, describe the correlated electronic wave function using a singlet unitary coupled-cluster single and double (UCCSD) ansatz \citep{Taube2006}. This ansatz is generated using the OpenFermion library \citep{openfermion} and then optimised using a VQE approach, where we use the modified Powell optimisation algorithm \citep{powell1964, numericalrecipes} implemented in the \textsc{SciPy} python \citep{2020SciPy-NMeth} library. We perform this first set of noisy simulations using a generic noise model implemented in IBM's QasmSimulator from the Qiskit library \citep{shortQiskit}. 

A second, more realistic set of noisy simulations is performed using IBM's QasmSimulator along with a detailed model of 7-qubit IBM Jakarta (\verb'ibmq_jakarta') device, accounting for coupling maps, basis gates, specific qubit noise and error. The noise profile from \verb'ibmq_jakarta' was collected on 20~June~2022 at 19:21:52~UTC and kept unchanged throughout the simulations. We mitigate error in the simulated values using a standard readout correction. 

In order to be able to run the calculations on a 7-qubit machine model, we adapt the approach of the first set of calculations as follows. First, we implement a Bravyi-Kitaev version of parity mapping which removes two qubits \citep{bravyi2017tapering}, requiring to 2 qubit for the MINI basis set calculations and 6 qubits for the 3-21G basis set calculations. Second, we use a hardware-efficient ansatz described in \cite{benfenati:2021}, instead of UCCSD, to reduce as much as possible the depth of the ansatz circuit. Third, we also introduce a simulation of the Hartree--Fock state as a means to mitigate measurement noise further. Details of both hardware ansatz specifications and related HF simulation are discussed in section \ref{sec:noisy}. 

All expectation values used for ansatz optimisations through the VQE procedure are obtained through averaging of 8192 samples (shots). 

\subsection{Result central tendency and variance estimation}
In order to capture a suitable central tendency for our VQE experiments, we repeat each experiment 8 times and use the mode or the median of the distribution instead of the mean. Indeed, since we are using 8 duplicates for each inter-nuclear separation, $r$, this technically is only a small statistical ensemble and in such cases, \cite{Dean:1951} recommend not to rely on the mean for central tendency. Both median and mode have the advantage of being resilient to the presence of outliers and for skewed statistics. Moreover, in a symmetric (normal) distribution the mode and median correspond to the mean. 

For the first set of noisy measurements, we compute each sample mode using the half-sample mode approach which provides a robust estimate even for a small sample. The variance of the computed mode is estimated using a Gaussian-kernel density estimation (KDE) smoothed bootstrap approach, following the suggestions of \citet{romano:88} and \citet{hedges:2003}. We typically use 10,000 mode estimations from our bootstrap sample and compute the 16-84 percentiles as proxy for a $1\sigma$ deviation.

We also explore the simpler approach outlined by \cite{Dean:1951} for our second set of more realistic noisy measurements that include a NISQ model. In this approach, we use the median only as a measure of central tendency and the standard deviation is estimated by a suitably weighted range ($\sigma=0.35\cdot \mathrm{range}$ for 8 observations). 

We see in practice that both mode/boostrap and median/range approaches lead to similar results, with the second technique being slightly less computationally demanding.

\section{Generic noise simulations}
\label{sec:generic}
\subsection{Potential energy curves}
\label{sec:curves}
The VQE results obtained for the potential energy curve of H$_2$ using the smallest basis set (MINI) and the largest basis set (3-21G) are shown in Fig.~\ref{fig:h2mini321g} (numerical values are available in the supplementary information section). We also report the results of the classical FCI calculations as reference. 

\begin{figure}[H]
 \centering
 \includegraphics[width=0.8\columnwidth]{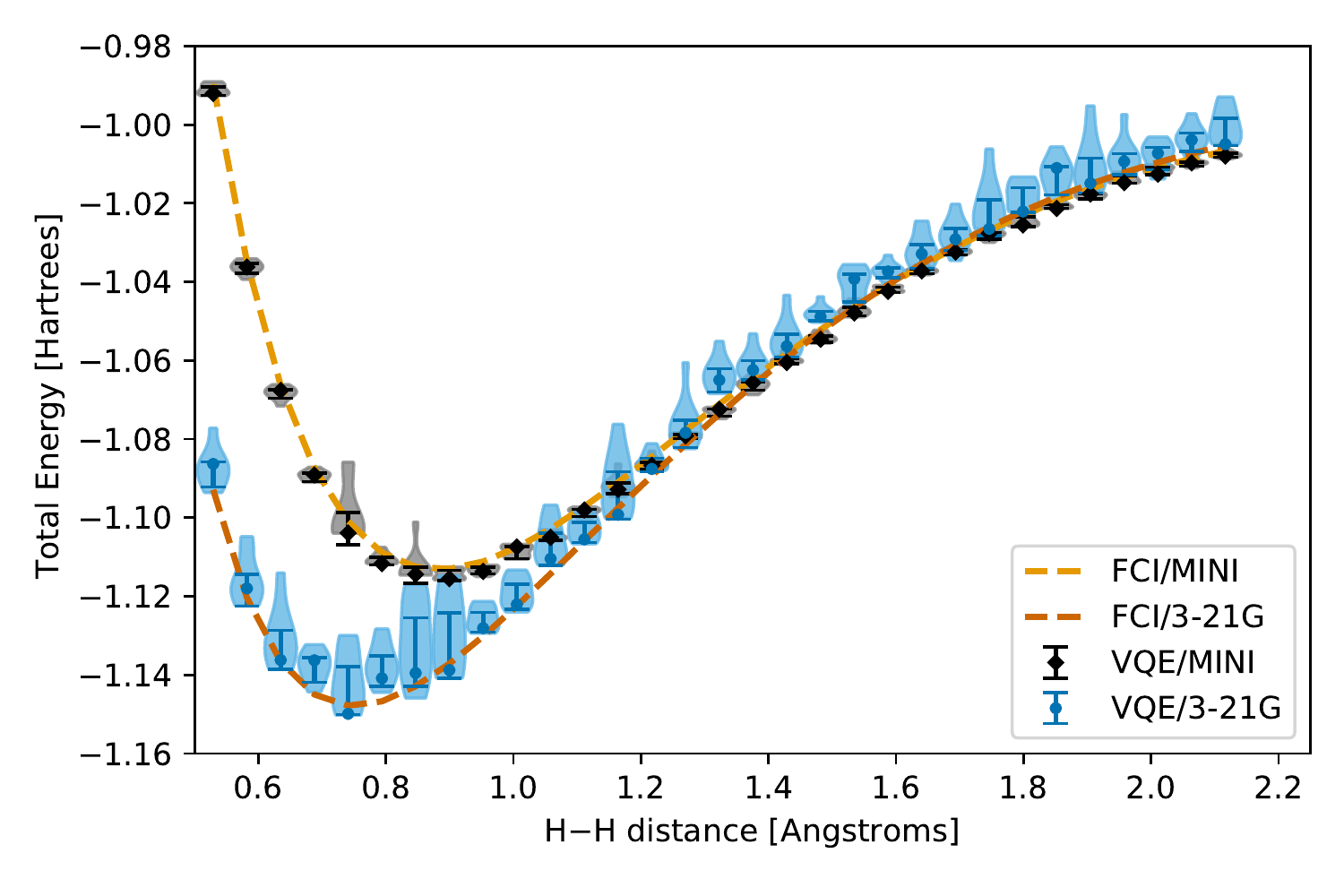}
 \caption{Potential energy curves for the H$_2$ molecule as a function of H--H separation obtained through VQE simulated experiments for the MINI (diamond) and 3-21G (circles) basis sets. Classically computed curves at the FCI level are also provided for each basis set. The shaded violin plots show the distribution of VQE results for each distance/basis set combination (gray shading: MINI, blue shading: 3-21G). The symbol indicates the mode for each series of experiment, along with a bootstrapped 16-84 percentile estimation of 1 standard deviation (see text for details).}
 \label{fig:h2mini321g}
\end{figure}

We see that within $1\sigma$ confidence limits, the results from the VQE calculations follow the reference curve well. For the smaller MINI basis, the points show enough differentiation between the energies at each distance that they define a curve suitable for further use in vibrational calculations, molecular dynamics or geometry optimisations, for example.

We note however a significant scatter of the VQE results at each inter-nuclear distance, which is more significant for the larger 3-21G basis set. This is likely caused by the larger number of qubits needed for this basis set and the resulting increase in depth and complexity of optimising its UCCSD ansatz. 

\begin{figure}[H]
 \centering
 \includegraphics[width=0.9\columnwidth]{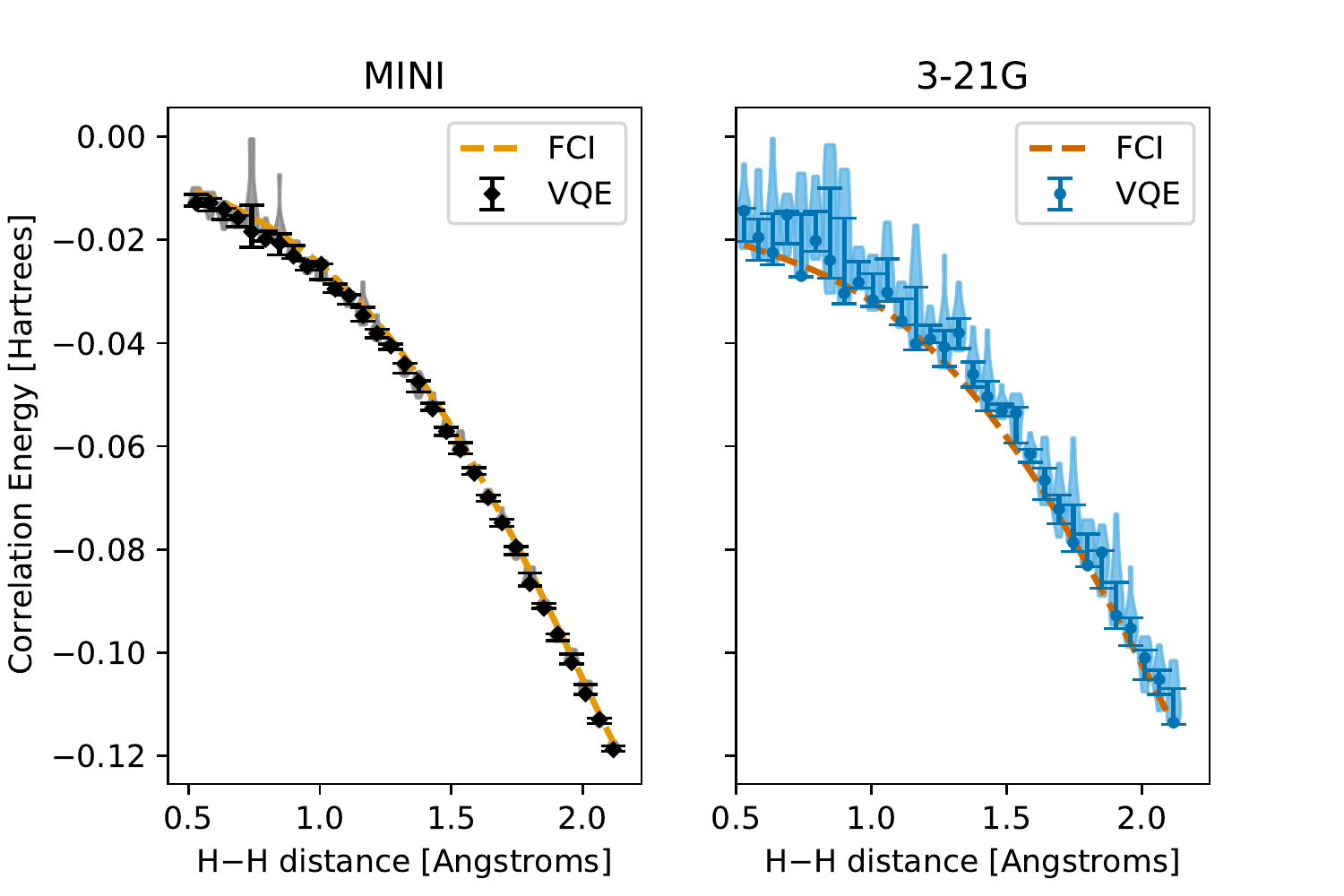}
 \caption{Electronic correlation energy curve for the H$_2$ molecule as a function of H--H separation obtained through VQE simulated experiments for the MINI (diamond) and 3-21G (circles) basis sets. Classically computed curves at the FCI level are also provided for each basis set. The shaded violin plots show the distribution of VQE correlation energies for each distance/basis set combination (gray shading: MINI, blue shading: 3-21G). The symbol indicates the mode for each series of experiment, along with a bootstrapped 16-84 percentile estimation of 1 standard deviation (see text for details).}
 \label{fig:h2mini321gcorr}
\end{figure}

The total energies shown in Fig.~\ref{fig:h2mini321g} all contain a significant proportion of Hartree--Fock energy, which is computed classically. A more stringent test of suitability is to compare the correlation energy component of the total energy obtained through VQE with the reference correlation at FCI level computed classically. Those results are shown in Fig.~\ref{fig:h2mini321gcorr}.

We see, here again, that the VQE results follow closely the reference curves for each basis set, albeit with more scatter for the larger 3-21G basis set. Nevertheless, we see that the agreement with the exact curve is good within a $1\sigma$ confidence interval.

\subsection{Hartree--Fock energy curve}
\label{sec:hf_curve}
\begin{figure}[H]
 \centering
 \includegraphics[width=\columnwidth]{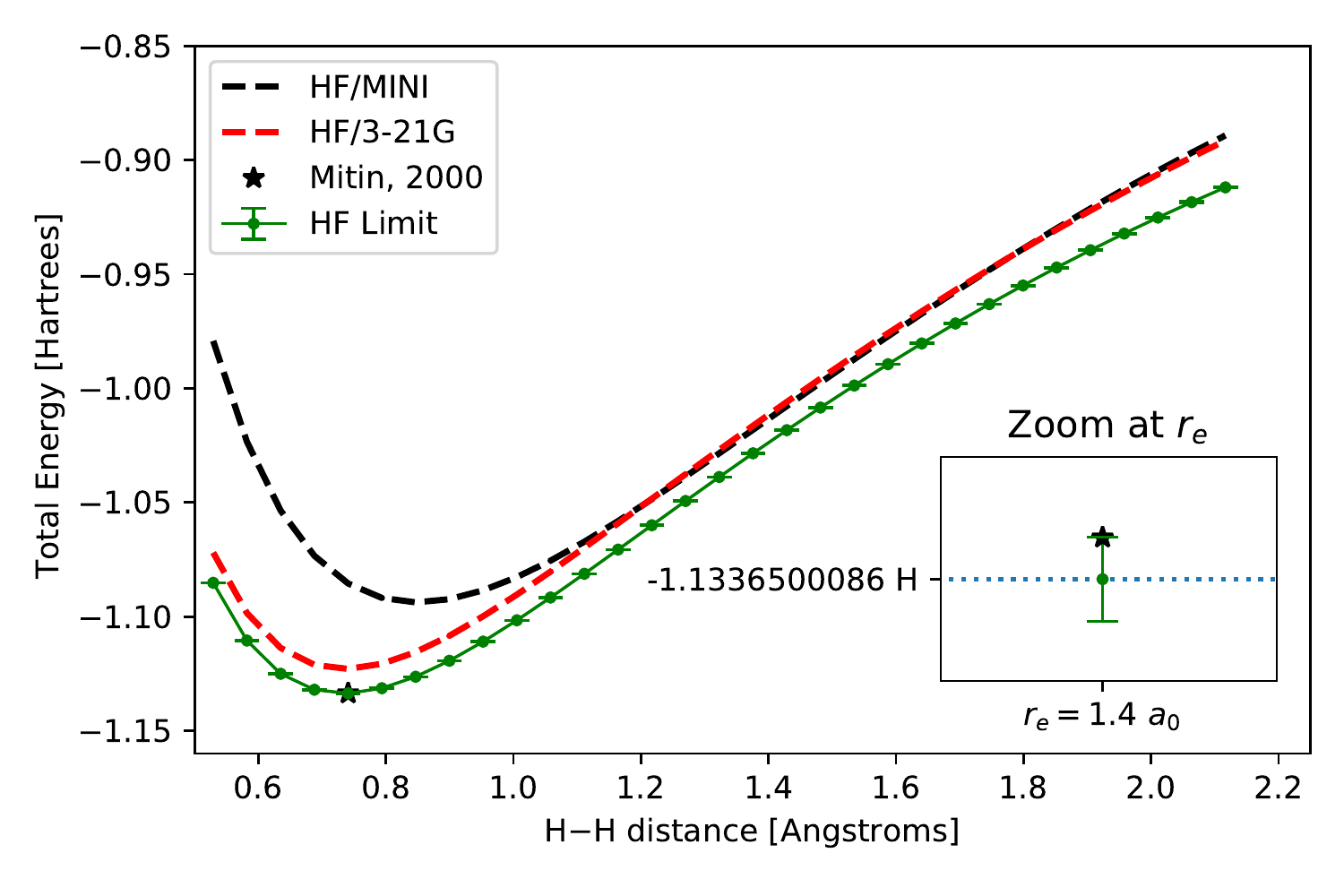}
 \caption{Hartree--Fock (HF) energy curve computed for H$_2$ extrapolated to the basis set limit using Jensen's pc-1, pc-2, pc-3 and pc-4 basis sets and Eq.~\ref{eq:hfenergy}. The classical HF curves computed with both MINI and 3-21G basis sets are shown for reference. We also include the HF limit value computed by \citet{mitin2000} using a 76-term basis set at the equilibrium radius for H$_2$ ($r_e=1.4$~bohrs or 0.7408481~{\AA}). Inset shows that our extrapolated curve agrees with the available reference data within error bars.}
 \label{fig:hflimit}
\end{figure}

Following Eq.~\ref{eq:totenergy}, the potential energy curve at the basis set limit can be constructed by combining a Hartree--Fock (HF) curve obtained for the basis set limit with the extrapolated correlation energy from the two FCI calculations. The basis set limit HF curve (obtained classically using Eq.~\ref{eq:hfenergy}) for H$_2$ is shown in Figure~\ref{fig:hflimit}, along with HF curves obtained with the MINI and 3-21G basis sets. We estimate the error bars for the extrapolated HF values using the standard deviation error from the fit of expression \ref{eq:hfenergy} to the computed HF data. 

Overall the mean error for our extrapolated HF results is 57~$\mu H$ (or 12.5~cm$^{-1}$) over the whole curve. The largest error occurs in the dissociation part of the curve, above 1.5~{\AA}, (up to 232~$\mu H$) but the estimated error in the binding region (0.6 -- 1.0~{\AA}) remains lower than 35~$\mu H$ (7.6~cm$^{-1}$). We note that our HF basis-set limit curve is a huge improvement over the MINI HF curve and a noticeable improvement over the 3-21G HF curve. Moreover, our HF results are in excellent agreement with the reference value of \citet{mitin2000}, obtained with a 76-term basis set in a variational calculation (within 1$\sigma$, see inset on right of Fig.~\ref{fig:hflimit}). However, all curves lead to a nonphysical dissociation limit due to the lack of explicit electronic correlation of restricted Hartree--Fock. 

\subsection{Electronic correlation curve}
\label{sec:corrcurve}
\begin{figure}[H]
 \centering
 \includegraphics[width=\columnwidth]{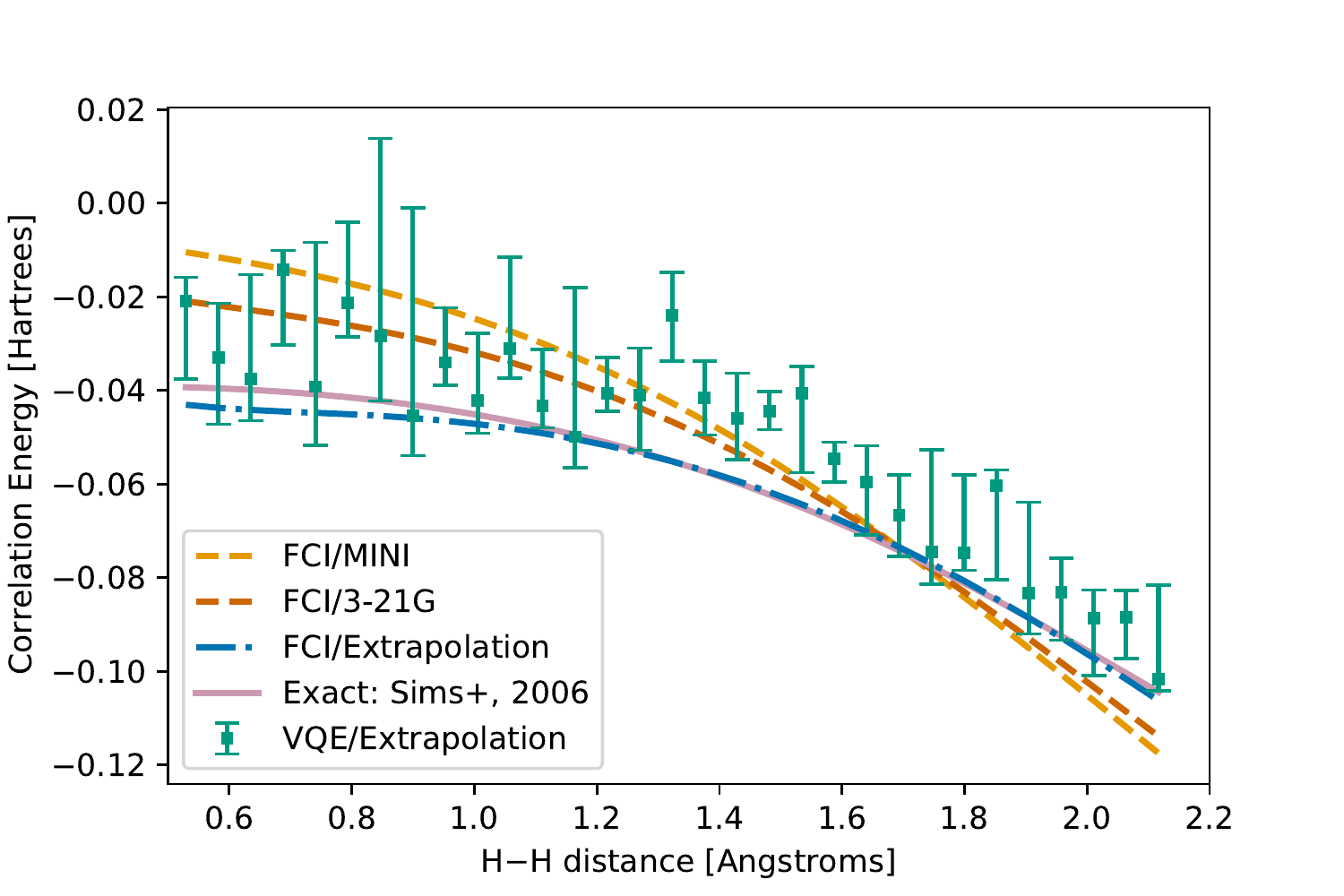}
 \caption{Electronic correlation energy curve for the H$_2$ molecule as a function of H--H separation. The extrapolated correlation energy is obtained through Eq.~\ref{eq:correnergy} (square symbols), based on the VQE correlation energies obtained using MINI and 3-21G basis sets (see also Fig.~\ref{fig:h2mini321gcorr}). Classically computed correlation energies at the FCI level are also provided for both MINI and 3-21G basis sets and extrapolation thereof, along with the "exact" correlation energy derived from the data of \citet{sims:2006} (purple line). The symbol indicates the mode for each extrapolation and the error bars indicate a Monte-Carlo 16-84 percentile estimation of 1 standard deviation (see text for details).}
 \label{fig:extrapcorr}
\end{figure}

In order to determine the second component of Eq.~\ref{eq:totenergy}, namely $E^\textrm{corr}_\infty(r)$, we use the VQE values obtained earlier to compute the correlation energy from each MINI and 3-21G basis sets (see also Fig.~\ref{fig:h2mini321gcorr}) and combine those using Eq.~\ref{eq:correnergy} to obtain an estimate of the correlation energy at the basis set limit. Our estimated correlation energy is shown in Fig.~\ref{fig:extrapcorr}, along with the "exact" correlation energy obtained by subtracting $E^\textrm{HF}_\infty(r)$ from the fully correlated curve of \citet{sims:2006}. Note that the latter curve is currently one of the most accurate fully-correlated Born-Oppenheimer data for H$_2$, with only that of \citet{Pachucki:2010} providing a slightly higher level of accuracy. Indeed, the \citet{sims:2006} results were obtained using a 7034-term wave function in a fully variational calculation. The approach used by \citet{Pachucki:2010} is similar but uses a larger 22,363-term wave function. These latter results provide a slight improvement over the \citet{sims:2006} curve but only on the scale of 10$^{-12}$ Hartrees, which is well below chemical accuracy.
 
First of all, we see from Fig.~\ref{fig:extrapcorr} that the extrapolation procedure performed on the classical FCI correlation data (dash-dotted blue line) leads to correlation estimates that are very close to the exact values (purple line). Indeed, the extrapolation improves on both MINI and 3-21G description by correctly reproducing the change in correlation situation as the molecule dissociates (after around 1.6 Angstroms).

We also see that our VQE estimates mirror the classical data, but show significant deviation in places, mainly due to the correlation energy estimates obtained for the larger 3-21G basis set. At this stage it is useful to keep in mind that we use a simple UCCSD ansatz, with little to no optimisation of either qubit or gate count in our calculations and thus improved estimator techniques are likely to lead to results that are closer to the classical estimates.

\citet{Varandas2018}, in his paper, provides three of values for $x_i$ for a given basis set: an MP2 value, a CCSD(T) value and an average value. In this work, we focus mainly on high-level correlation methods (i.e.\ FCI) which is not technically described in the paper (albeit CCSD(T) admittedly can be close to FCI for some systems). Therefore, it seems appropriate to use the average values suggested by \citet{Varandas2018} ($\langle x_\textrm{MINI}\rangle =1.450$ and $\langle x_\textrm{3-21G}\rangle=1.637$), as they cover a more generic correlation situation and provide a robust extrapolation by removing one source of error. 

We observe in Fig.~\ref{fig:h2mini321gcorr} that the data follows an asymmetric distribution which is quite marked for the 3-21G VQE experiments and slightly less so for the VQE MINI experiments. This non-normality can create further complications for error/variance determination, as most statistical approaches usually assume a symmetrical error distribution.
 
An alternative estimation of the variability for the extrapolated values can also be obtained by performing a Monte-Carlo (MC) simulation of the data distribution for $E^\textrm{corr}_{x_i}$ and gather statistics on the outcome of Eq.~\ref{eq:correnergy}. Essentially, this corresponds to exploring the variability space of the extrapolated data, given a known variability of the VQE results. The resulting MC data can then be used to construct a 16-84 percentile interval which is comparable to the $1\sigma$ deviation commonly used to describe variability for normal distribution (example python code provided in supplementary information). 

Note that the Monte-Carlo procedure is typically very stable and depends only weakly on the simulation conditions used. In this work, we use a random draw of 10,000 random variates, sampled from the distributions of errors obtained from the previous section. The error bars obtained are large but likely provide a realistic estimate of the true error on the extrapolated correlation values. 

\subsection{Full energy curve}
\label{sec:full_curve}
\begin{figure}[H]
 \centering
 \includegraphics[width=\columnwidth]{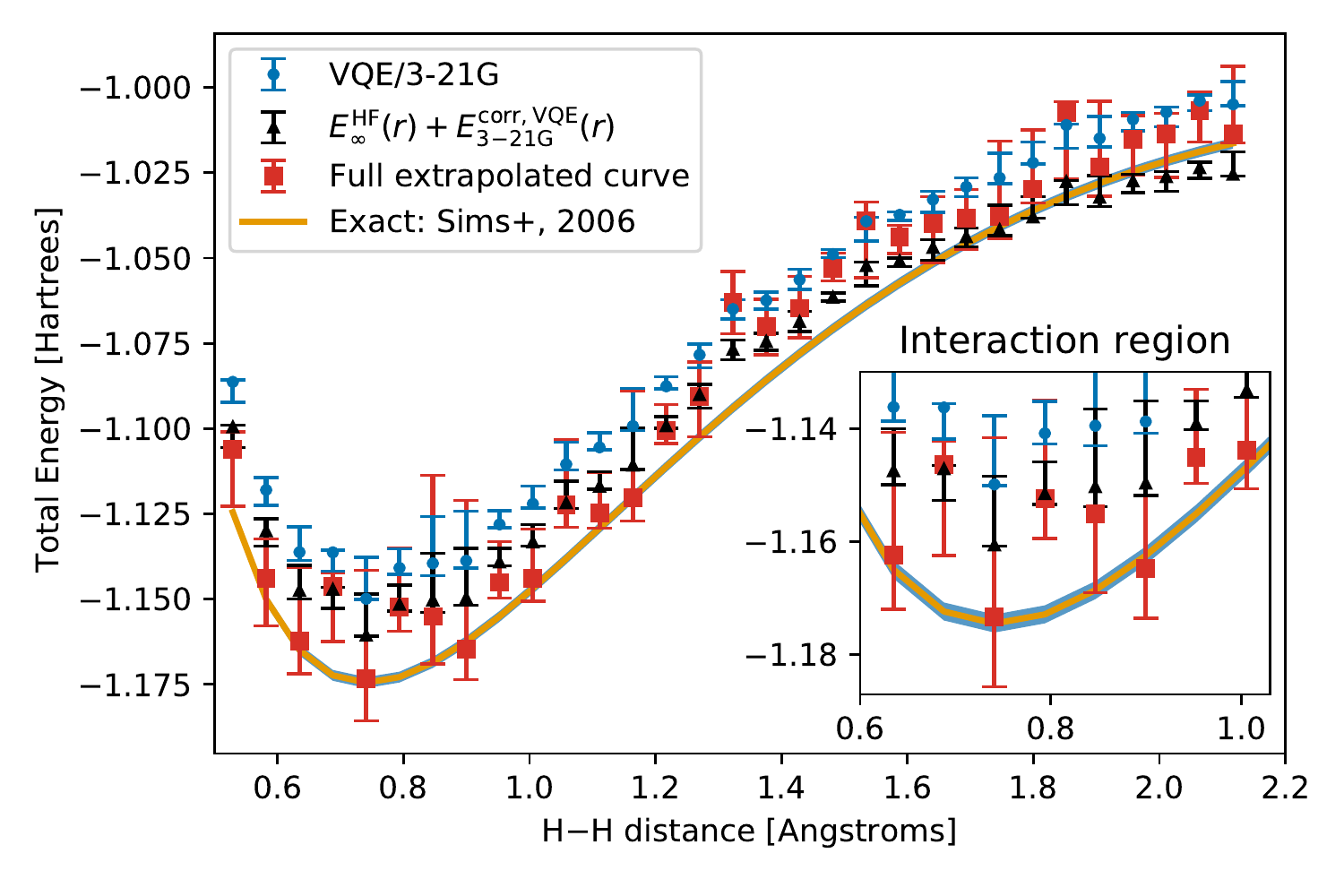}
 \caption{Total electronic energy of H$_2$ as a function of H--H separation. The 3-21G VQE estimations are shown as blue circles, an alternative approximation using $E^\textrm{HF}_\infty(r)$ and the 3-21G VQE correlation energy is shown as black triangles. The full extrapolated curve, using both $E^\textrm{HF}_\infty(r)$ and $E^\textrm{corr}_\infty$, is shown as red squares. Finally, the "exact" energy curve obtained classically by \citet{sims:2006} is shown as an orange continuous line, with an indication of the chemical accuracy limit (1.6~mH) shown as a blue outline around the curve. The symbols indicate the mode for set of data points and the error bars indicate a bootstrap/Monte-Carlo 16-84 percentile confidence interval (see text for details). Inset highlights the interaction region where we see that our extrapolated curve improves over the other approximations (VQE only or $E^\textrm{HF}_\infty(r)+E^\textrm{corr,VQE}_\mathrm{3-21G}(r)$ correlation) and in places fully matches the available reference data.}
 \label{fig:full_curve}
\end{figure}

By combining the results obtained from the extrapolated HF curve (see also Sec.~\ref{sec:hf_curve}) with the electronic correlation estimates from the previous section (Sec.~\ref{sec:corrcurve}), we can obtain an estimate of the total potential energy curve for H$_2$. The resulting curve is shown in Fig.~\ref{fig:full_curve}, along with the reference curve of \citet{sims:2006}. 

Here again, since the correlation error is intrinsically asymmetric, we cannot use a simple combination rule for the two errors (HF and correlation error) to obtain the error on our combined curve. The unsuitability of a simple quadratic error formula for asymmetric errors has been extensively discussed in \citet{laursen_error:2019} and \citet{Barlow:2003}. Here, we use the approximate procedure of \citet{Barlow:2003} which has been implemented in python by the lead authors of \citet{laursen_error:2019}\footnote{freely available at: \href{https://github.com/anisotropela/add_asym}{https://github.com/anisotropela/add\_asym}}. 

Overall, we see that the correlation estimates essentially dominate the quality of the curves obtained. We observe relatively small error bars for the 3-21G VQE total energy calculations, but this basis set produces results that are far from the exact FCI values. A slightly better approximation is to extract the correlation energy from the 3-21G VQE experiments, $E^\textrm{corr,VQE}_\mathrm{3-21G}(r)$, and replace the 3-21G HF contribution with the extrapolated classical HF values, $E^\textrm{HF}_\infty(r)$. Indeed, this bring the resulting curve closer to the reference data, since the extrapolated HF curve is a notable improvement over the 3-21G HF data (see Fig.~\ref{fig:hflimit}) and keeps VQE correlation error minimal. However, this approximation still displays a number of issues: it underestimates the energy in the interaction region and over-estimates the values towards dissociation as the representation of electronic correlation by a 3-21G basis alone is sub-optimal (as shown in Fig.~\ref{fig:extrapcorr} for the classical FCI/3-21G values).

Finally, combining the fully extrapolated correlation energies, $E^\textrm{corr,VQE}_\infty(r)$, with $E^\textrm{HF}_\infty(r)$ admittedly leads to a noisier dissociation curve, but we see that within error bars, those results agree very well with the reference curve. We simultaneously improve the agreement in the interaction region but also fix the dissociation limit - overall leading to a more realistic curve. As stated earlier, this approach is still dominated by the uncertainty originating from two sets of VQE experiments, but essentially, this shows that NISQ devices can theoretically approach the full correlation limit and full basis set limit \emph{without} the need for a large number of qubits. Indeed, the required number of qubits remains very modest and less noisy implementations of the VQE approach would further improve the quality of the agreement. 

If we focus on the interaction region only, between 0.6~{\AA} and 1.0~{\AA}, we find that the root mean square deviation from the exact curve is 13~mH for our fully extrapolated curve, improving on 18~mH for $E^\textrm{HF}_\infty(r)+E^\textrm{corr,VQE}_\mathrm{3-21G}(r)$ and 28~mH for the 3-21G VQE results. Keeping in mind that chemical accuracy (1~kcal/mol) is at around 1.6~mH, we see that this level of accuracy would require an order of magnitude improvement in the interaction region. Further developments in this direction are already underway in our laboratory. 

\section{Realistic noise simulation data}
\label{sec:noisy}
In order to evaluate the usefulness of our approach on actual NISQ devices, we performed the same calculations but using a machine-specific noise model of \verb'ibmq_jakarta' that accounts for coupling maps, basis gates, specific qubit noise and error.

\subsection{Choice of ansatz}
\begin{figure}[h!]
 \centering
\scalebox{1.0}{
\Qcircuit @C=1.0em @R=0.2em @!R {
 & & & & & &&\mbox{Repeat $n_\ell$ times} \\
	 	\nghost{{q}_{0} : } & \lstick{{q}_{0} : } & \gate{\mathrm{R_Y}\,(\mathrm{{\ensuremath{\theta}}[0]})} & \qw & \ctrl{1} & \qw & \qw & \qw & \qw &\qw & \gate{\mathrm{R_Y}\,(\mathrm{{\ensuremath{\theta}}[6]})} & \qw&\qw\\
	 	\nghost{{q}_{1} : } & \lstick{{q}_{1} : } & \gate{\mathrm{R_Y}\,(\mathrm{{\ensuremath{\theta}}[1]})} & \qw & \targ & \ctrl{1} & \qw & \qw & \qw & \qw &\gate{\mathrm{R_Y}\,(\mathrm{{\ensuremath{\theta}}[7]})} & \qw & \qw\\
	 	\nghost{{q}_{2} : } & \lstick{{q}_{2} : } & \gate{\mathrm{R_Y}\,(\mathrm{{\ensuremath{\theta}}[2]})} & \qw & \qw & \targ & \ctrl{1} & \qw & \qw & \qw &\gate{\mathrm{R_Y}\,(\mathrm{{\ensuremath{\theta}}[8]})} & \qw & \qw\\
	 	\nghost{{q}_{3} : } & \lstick{{q}_{3} : } & \gate{\mathrm{R_Y}\,(\mathrm{{\ensuremath{\theta}}[3]})} & \qw & \qw & \qw & \targ & \ctrl{1} & \qw & \qw &\gate{\mathrm{R_Y}\,(\mathrm{{\ensuremath{\theta}}[9]})} & \qw & \qw\\
	 	\nghost{{q}_{4} : } & \lstick{{q}_{4} : } & \gate{\mathrm{R_Y}\,(\mathrm{{\ensuremath{\theta}}[4]})} & \qw & \qw & \qw & \qw & \targ & \ctrl{1} & \qw &\gate{\mathrm{R_Y}\,(\mathrm{{\ensuremath{\theta}}[10]})} & \qw & \qw\\
 	 	\nghost{{q}_{5} : } & \lstick{{q}_{5} : } & \gate{\mathrm{R_Y}\,(\mathrm{{\ensuremath{\theta}}[5]})} & \qw & \qw & \qw & \qw & \qw & \targ & \qw &\gate{\mathrm{R_Y}\,(\mathrm{{\ensuremath{\theta}}[11]})} & \qw & \qw\\
 {\gategroup{2}{4}{7}{11}{.8em}{--}} \\
 }}
 \caption{Example of a 6-qubit $n_\ell$-layer hardware-efficient ansatz circuit, based on $\mathrm{R_Y}$ gates and linear $\mathrm{C_{NOT}}$ gates. Here we show a single layer ($n_\ell=1$) used to generate a 12-parameter variational form for our VQE experiments that include machine-derived noise models. Note that each additional layer adds another 6 $\mathrm{R_Y}$ parameters.}
 \label{fig:francesco_ansatz}
\end{figure}
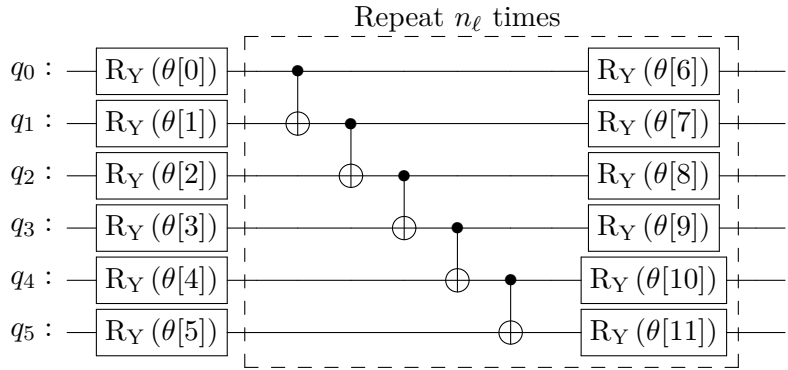

Initial tests showed that the UCCSD ansatz used in the previous section is often not optimised efficiently under realistic noise conditions. Instead, we use a heuristic $n_\ell$-layer $R_Y-\{C_{NOT}-R_Y\}_{n_ \ell}$ hardware-efficient ansatz (HEA) for those calculations, as described in Fig.~1 of \cite{benfenati:2021}. An example of such ansatz for the 3-21G basis set calculation using a BK-parity mapping (6 qubits) is shown in Fig.\ref{fig:francesco_ansatz}. 

\begin{figure}[H]
 \centering
 \includegraphics[width=0.6\columnwidth]{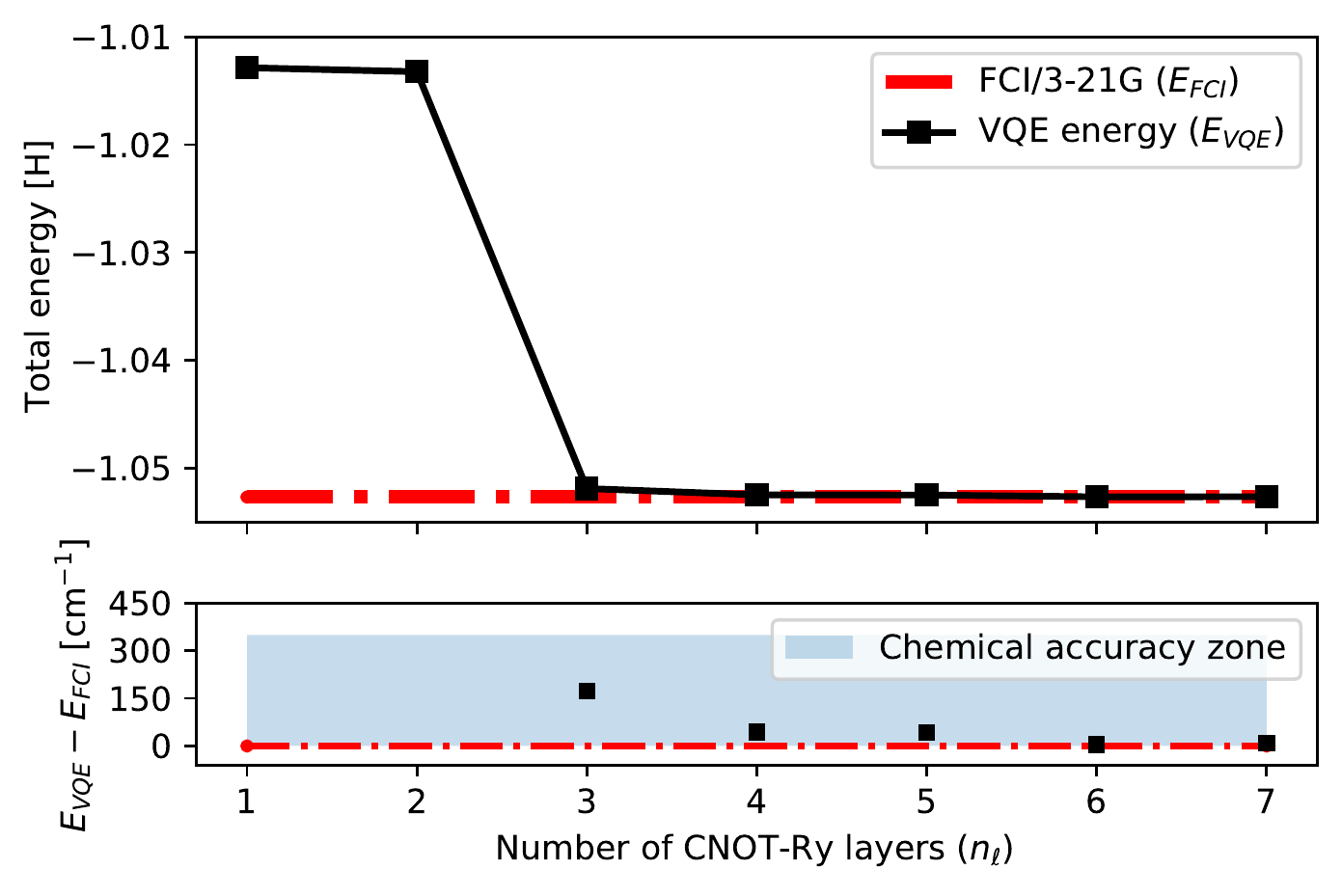}
 \caption{Convergence behaviour of the hardware-efficient ansatz (HEA[$n_\ell$]) as a function of the number of layers ($n_\ell$) optimised using the BFGS algorithm. The total energy (shown in Hartrees) is computed using a state-vector (noiseless) simulation of the H$_2$ system at $r=1.481696$~{\AA}. The electronic wave function is described using a 3-21G basis with BK-parity mapping. The lower part of the graph shows a detailed view of the convergence behaviour (shown in cm$^{-1}$), along with a shaded area indicating chemical accuracy ($<1$~kcal/mol).}
 \label{fig:HEAlayers}
\end{figure}

Such ansatz has been used successfully in many studies, but does not always allow enough flexibility to reach the FCI reference values with a single layer (see below). However, increasing the number of layers in the ansatz leads to a good convergence to the exact FCI value with a BFGS optimiser, for example. We show in Fig.~\ref{fig:HEAlayers} a typical convergence profile (computed at $r=1.481696$~{\AA}) for this HEA as a function of $n_\ell$ for a noiseless simulation of the 3-21G basis with BK-parity mapping. We observe that the ansatz is close to convergence for $n_\ell=3$ already, with the largest improvement obtained by going from $n_\ell=2$ to $n_\ell=3$. In particular, we see that this ansatz reaches the chemical accuracy zone for $n_\ell\geq 3$ (blue zone in the lower panel of Fig.~\ref{fig:HEAlayers}). In order to simplify notation, we will use the notation HEA[$n_\ell$] to signify an $n_\ell$-layer HEA ansatz. 

\subsection{Noise mitigation using a mean-field ground state}
\begin{figure}[H]
 \centering
 \includegraphics[width=0.8\columnwidth]{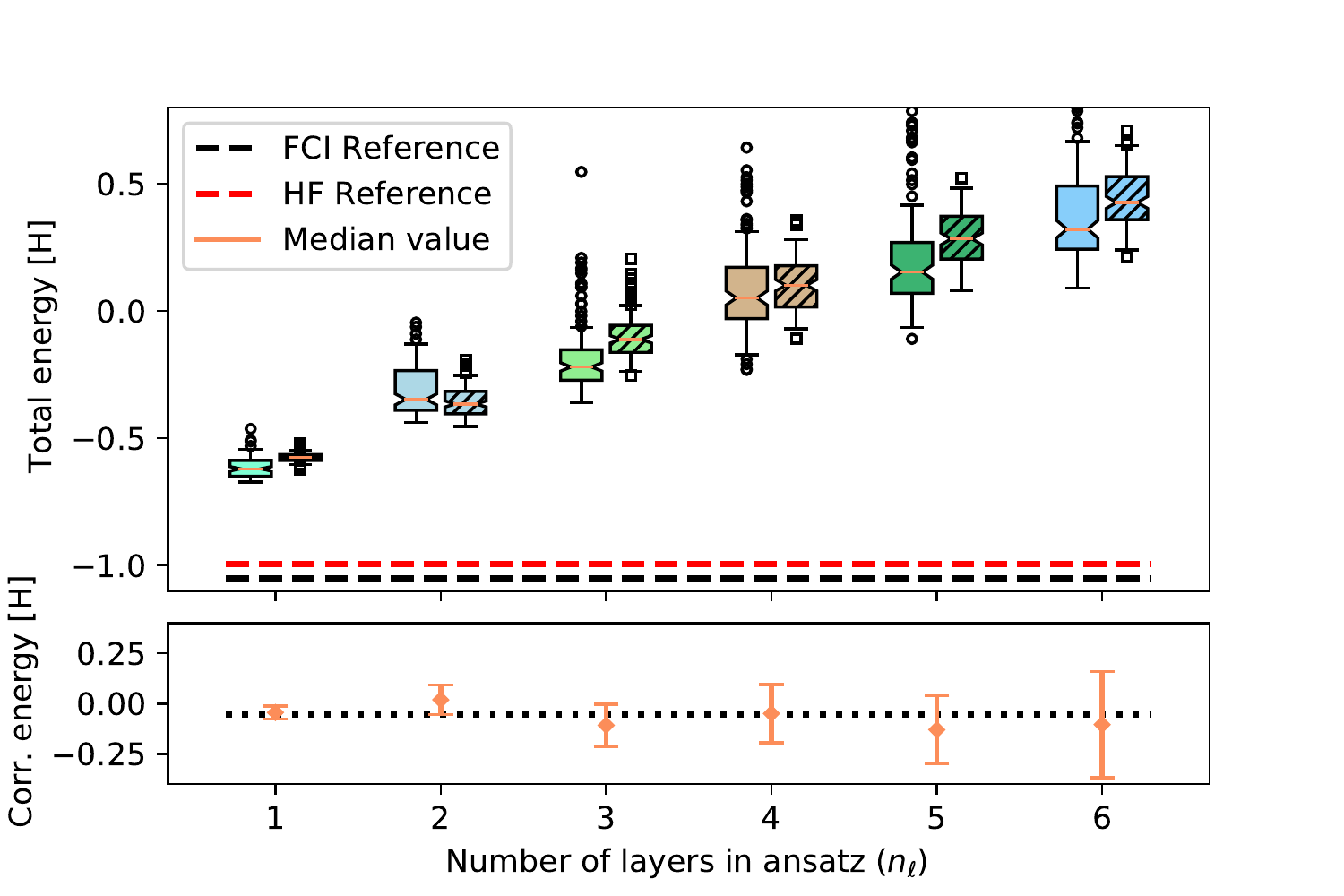}
 \caption{Realistic-noise simulation for the total energy of the H$_2$ molecule (3-21G basis set and BK-parity mapping). The H--H distance is held fixed at $r=1.481696$~{\AA} and each measurement is repeated 128 times for increasing number of layers $n_\ell$. In the top panel, the left (unhatched) boxplots show results for a set of HEA[$n_\ell$] optimisations, while the right boxplots (hatched) show the results for a set of HFHEA[$n_\ell$] optimisations. The median of each data set is shown as a notch and orange line in the box plots, along with the FCI reference energy (black dashed line) and the HF reference line (red dashed line). Note that the y-axis on the graph has been clipped for visualisation purposes as both $n_\ell=5$ and $n_\ell=6$ experiments have a range that extends to $+0.78552766$~H and $+1.7841259$~H, respectively. The bottom panel shows the median of the estimated correlation energy ($E_\mathrm{FCI}-E_\mathrm{HF}$) for each ansatz depth, along with the reference value for the 3-21G basis set at this inter-nuclear separation (dotted black line). The error bars indicate 1 standard deviation ($1\sigma$).}
 \label{fig:rawenergies-321g-noisy}
\end{figure}

The realistic noise simulations cause a significant drift in the computed total energy of the simulated H$_2$ molecule. A typical example for the values obtained with the 3-21G basis set and a BK-parity mapping is shown in Fig.~\ref{fig:rawenergies-321g-noisy} (top panel, left box plots) for $r=1.481696$~{\AA}. We notice that, not only does the median total energy estimate increases as the number of layers increases, but the total energy values obtained are significantly higher than both the FCI reference value and the Hartree--Fock reference value for this basis set. This would usually lead to nonphysically large and \emph{positive} correlation energy values, which cannot be used. We show in the following section how we can nevertheless mitigate this error using a quantum circuit that evaluates the HF ground state energy.

\subsubsection{Equivalent mean-field ansatz}
If we consider the results shown in the top panel of Fig.~\ref{fig:rawenergies-321g-noisy}, we observe two effects. First, the total energy increases nearly linearly as a function of $n_\ell$. Second, we see an increased statistical spread of the data obtained as $n_\ell$ increases ($n_\ell=1$ values show much less scatter than $n_\ell=6$, for example). We can assume that that both shift and scatter originate from hardware noise (gate errors, qubit noise, connectivity limitations) and possible errors and/or optimisation bias in the VQE procedure. Thus, we postulate the following simple model for our noisy measurements:
\begin{equation}
E_{VQE}^\mathrm{meas.}=E_{}^\mathrm{true}+\epsilon_e
 \label{eq:noisedef}
\end{equation}
where $\epsilon_e$ is a random error and we assume that $\epsilon_e\sim \mathcal{N}(\mu_e,\sigma_e^2)$, i.e.\ the error is distributed normally (in line with standard error theory, see for example \cite{errorbook}). In this model, the distribution of $\epsilon_e$ accounts for both energy shifting through its parameter $\mu_e$ and scattering of data through $\sigma_e$. We take $E_{}^\mathrm{true}$ to be the true value of the energy the VQE procedure is trying to estimate and assume that the distribution of $E_{VQE}^\mathrm{meas.}$ is accessible through a statistical analysis of the VQE experiments. These assumptions are reasonable but the normality of the data is not necessarily guaranteed in practice (in particular, we have some indications in Fig.~\ref{fig:h2mini321g} and Fig.~\ref{fig:rawenergies-321g-noisy} that the distribution is possibly non-normal for large circuit depth). 

In order to characterise parameters that define the error distribution, $\mathcal{N}(\mu_e,\sigma_e)$, we perform a set of VQE estimations of a known quantity: the Hartree--Fock (HF) ground state. Indeed, as the HF reference value is known for each inter-nuclear distance ($r$) from the construction of the second-quantisation Hamiltonian, we can use Eq.~\ref{eq:noisedef}, re-written as:
\begin{equation}
 \epsilon_e(r)=E_{VQE-HF}^\mathrm{meas.}(r)-E_{HF}^\mathrm{true}(r)
 \label{eq:HFnoisedef}
\end{equation}
where we assume that $\epsilon_e$ varies with $r$ and hence that the value of $\mu_e$ and $\sigma_e$ that describe the error distribution are also distance-dependent. Here, $E_{VQE-HF}^\mathrm{meas.}(r)$ is the measured expectation values of the HF ground state and $E_{HF}^\mathrm{true}(r)$ is the exact HF energy at inter-nuclear separation $r$. 

\begin{figure}[h!]
 \centering
\scalebox{1.0}{
\Qcircuit @C=0.7em @R=0.3em @!R {
 & & &&&&&\mbox{Repeat $n_\ell$ times} \\
	 	\nghost{{q}_{0} : } & \lstick{{q}_{0} : } & \gate{\mathrm{X}}
 & \gate{\mathrm{R_Y}\,(\mathrm{{\ensuremath{+\theta}}[0]})} 
 & \gate{\mathrm{R_Y}\,(\mathrm{{\ensuremath{-\theta}}[0]})} 
 & \gate{\mathrm{X}}& \ctrl{1} &\gate{\mathrm{X}}& \qw & \qw & \qw & \qw & \qw & \qw & \qw & \qw& \qw\\
	 	\nghost{{q}_{1} : } & \lstick{{q}_{1} : } & \gate{\mathrm{X}}
 & \gate{\mathrm{R_Y}\,(\mathrm{{\ensuremath{+\theta}}[1]})}
 & \gate{\mathrm{R_Y}\,(\mathrm{{\ensuremath{-\theta}}[1]})}
 & \qw&\targ &\gate{\mathrm{X}}& \ctrl{1} & \gate{\mathrm{X}}&\qw & \qw & \qw & \qw& \qw& \qw& \qw\\
	 	\nghost{{q}_{2} : } & \lstick{{q}_{2} : } &\qw
 & \gate{\mathrm{R_Y}\,(\mathrm{{\ensuremath{+\theta}}[2]})}
 & \gate{\mathrm{R_Y}\,(\mathrm{{\ensuremath{-\theta}}[2]})}
 & \qw & \qw & \qw& \targ & \ctrl{1} & \qw & \qw & \qw & \qw& \qw& \qw& \qw\\
	 	\nghost{{q}_{3} : } & \lstick{{q}_{3} : } & \gate{\mathrm{X}}
 &\gate{\mathrm{R_Y}\,(\mathrm{{\ensuremath{+\theta}}[3]})} 
 &\gate{\mathrm{R_Y}\,(\mathrm{{\ensuremath{-\theta}}[3]})} 
 & \qw & \qw & \qw & \qw & \targ &\gate{\mathrm{X}}& \ctrl{1} & \gate{\mathrm{X}}&\qw& \qw& \qw& \qw \\
	 	\nghost{{q}_{4} : } & \lstick{{q}_{4} : } & \gate{\mathrm{X}}
 &\gate{\mathrm{R_Y}\,(\mathrm{{\ensuremath{+\theta}}[4]})} 
 &\gate{\mathrm{R_Y}\,(\mathrm{{\ensuremath{-\theta}}[4]})} 
 & \qw & \qw & \qw & \qw & \qw & \qw &\targ & \gate{\mathrm{X}}&\ctrl{1} & \gate{\mathrm{X}}&\qw& \qw\\
 	 	\nghost{{q}_{5} : } & \lstick{{q}_{5} : } & \qw
 &\gate{\mathrm{R_Y}\,(\mathrm{{\ensuremath{+\theta}}[5]})} 
 &\gate{\mathrm{R_Y}\,(\mathrm{{\ensuremath{-\theta}}[5]})}
 & \qw &\qw & \qw & \qw & \qw & \qw & \qw & \qw & \targ & \qw & \qw& \qw\\
 {\gategroup{2}{4}{7}{16}{.6em}{--}} \\
 }}
 \caption{Example of a Hartree--Fock 6-qubit ansatz (HFHEA[$n_\ell$]) with a structure compatible with the $n_\ell$-layer hardware-efficient ansatz circuit. This ansatz is designed to contain approximately as many gates as an HEA[$n_\ell$] but rotations are performed in a specific sequence ($+\theta$ followed by $-\theta$) to create an overall \emph{null} rotation and thus prevent any changes to the starting HF state during the VQE experiments. Here we show a single layer ($n_\ell=1$).}
 \label{fig:HF_ansatz}
\end{figure}
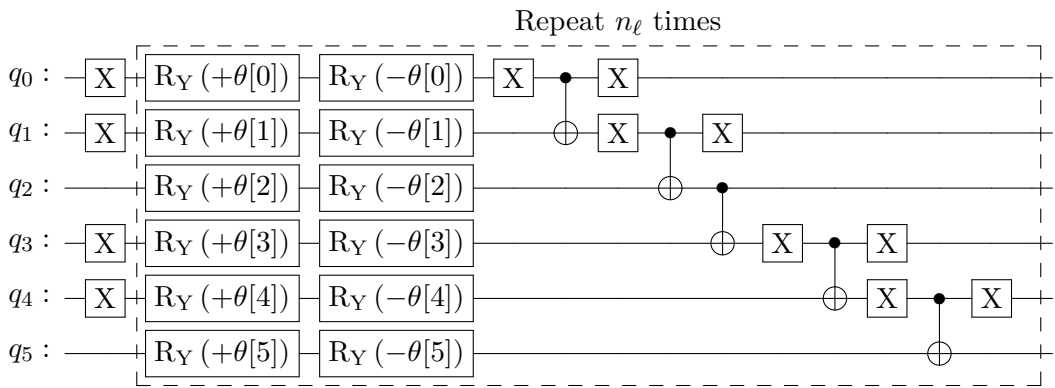

The sampling of the HF ground state expectation value is performed using a modified HEA[$n_\ell$] ansatz, named HFHEA[$n_\ell$] from here on, with a similar circuit depth to that of the original HEA[$n_\ell$]. A typical example of the HFHEA[$n_\ell$] for the BK-parity mapping with 6-qubit is shown in Fig.~\ref{fig:HF_ansatz}. 

The first set of X gates on the left of the ansatz prepare a HF grounds state for the BK-parity mapping. This prepared ground state is then combined with a sequence of two rotations ($+\theta$ followed by $-\theta$) mimicking the HEA[$n_\ell$] rotation gates but preventing any changes to the HF reference. We also modify the original CNOT gates required by the HEA[$n_\ell$] anzatz to X-CNOT-X gates for the occupied qubits to ensure that identity is preserved. The HFHEA[$n_\ell$] ansatz is then used in a VQE procedure to sample the fidelity of the state estimation, as the exact value is already known from the second quantisation Hamiltonian. Note however that since there is no optimisable degrees of freedom in this ansatz, we likely do not sample possible optimisation bias of the VQE procedure fully.

Typically, we perform an optimisation of the HF state just before each HEA[$n_\ell$] optimisation in order to ensure that both measurements are close enough in time. On an actual device, this would help minimise the effects of hardware re-calibration cycles if each pair of measurements is considered together. 

\subsubsection{Noisy correlation energy computation}
\label{sec:corrhfvqe}
Since the extrapolation procedure described in Sec.~\ref{sec:CBSth} requires the correlation energy at each distance, we need to recast Eq.~\ref{eq:correnergylevelx} to a noisy context. A naive approach is to simply subtract the exact (or true) HF energy from each FCI VQE results. Following the assumptions of equation \ref{eq:noisedef}, this can be written down as:
\begin{eqnarray}
E_\mathrm{VQE-corr}^\mathrm{meas.}&=&E_\mathrm{VQE-FCI}^\mathrm{meas.}-E_\mathrm{HF}^\mathrm{true}\nonumber\\
&=&\left({E_{FCI}^\mathrm{true}+\epsilon^\prime_e}\right)-E_\mathrm{HF}^\mathrm{true}\nonumber\\
&=&E_{FCI}^\mathrm{true}-E_\mathrm{HF}^\mathrm{true}+\epsilon^\prime_e\nonumber\\
&=&E_\mathrm{corr}^\mathrm{true}+\epsilon^\prime_e \label{eq:naivecorr}
\end{eqnarray}

This is similar to the expression derived earlier and leads to a noisy estimation of the true correlation energy. Here we denote the random error from the VQE estimation of the FCI ground state by $\epsilon^\prime_e$, as it likely also contains some optimisation bias. Unfortunately, our noisy simulations show that the magnitude of $\epsilon^\prime_e$ is often too large to afford a meaningful estimation of $E_\mathrm{VQE-corr}$ through this simple route (see for example Fig.~\ref{fig:rawenergies-321g-noisy}, where all FCI estimates are very far above the true HF energy).

An alternative approach is to use the noise sampled during the HF simulation as an internal standard to effectively "re-calibrate" our FCI results. Indeed, if we compute the correlation energy using $E_{HF}^\mathrm{meas.}$ instead, we obtain:
\begin{eqnarray}
E_\mathrm{corr}^\mathrm{meas.}&=&E_\mathrm{VQE-FCI}^\mathrm{meas.}-E_\mathrm{VQE-HF}^\mathrm{meas.}\nonumber\\
&=&\left(E_{FCI}^\mathrm{true}+\epsilon^\prime_e\right)-\left(E_{HF}^\mathrm{true}+\epsilon_e\right)\nonumber\\
&=&\left(E_{FCI}^\mathrm{true}-E_{HF}^\mathrm{true}\right)+\underbrace{\left(\epsilon^\prime_e-\epsilon_e\right)}_{\epsilon^{\prime\prime}_e}\nonumber\\
&=&E_\mathrm{corr}^\mathrm{true}+\epsilon^{\prime\prime}_e\label{eq:gausscorr}
\end{eqnarray}

where $\epsilon^{\prime\prime}_e=\epsilon^\prime_e-\epsilon_e$ is a combination of the two random errors for each measurements. We assume that the hardware-noise component of $\epsilon^\prime_e$ is described by $\epsilon_e$ and that the difference (i.e.\ $\epsilon^{\prime\prime}_e$) corresponds to any remaining VQE optimisation bias/error not sampled by our HFHEA[$n_\ell$] ansatz. If we combine this assumption with the normal sum theorem, we can show that $\epsilon^{\prime\prime}_e\sim\mathcal{N}(\mu^\prime_e-\mu_e,\sigma^\prime_e{}^2-\sigma_e^2)$, which performs an approximate error deconvolution. In practice, we use this approach to compute the correlation energy by subtracting the median value of $E_\mathrm{VQE-HF}^\mathrm{meas.}(r)$ from $E_\mathrm{VQE-FCI}^\mathrm{meas.}(r)$ and standard deviation through $\sigma_\mathrm{corr.}(r)=\sqrt{\sigma^\prime_e(r)^2-\sigma_e(r)^2}$. Note that this model assumes that all random errors follow a normal distribution. \\

While Eq.~\ref{eq:gausscorr} appears at first sight to be similar to Eq.~\ref{eq:naivecorr}, the present expression corrects both for the observed energy shift and hardware noise to some extent. An example of the performance of this approach is shown in the bottom panel of Fig.~\ref{fig:rawenergies-321g-noisy}, where we see that the exact correlation is recovered for all ansatz depths, within the error bars. We also see that the number of layers has a direct influence on the scatter of the correlation estimations. It is worth noting at this stage that the naive approach of Eq.~\ref{eq:naivecorr} leads to erroneous nonphysical correlation values, as can be inferred from the upper panel of Fig.~\ref{fig:rawenergies-321g-noisy} (distance between median of VQE data and black dashed line).

The use of a combined set of VQE measurements for both the HF ground state and the fully-correlated ground state, enables us to reject data points that lead to nonphysical positive correlation energy. This is implemented in our approach and in practice leads to 1\% rejections for the MINI basis results and 32\% for the data obtained with the larger 3-21G basis. While the latter rejection rate is high, it is not detrimental to the correlation energy estimation. 

\subsection{Choice of ansatz depth}
While a BFGS optimiser is an efficient optimiser for HEA[$n_\ell$] in noiseless simulations (see also Fig.~\ref{fig:HEAlayers}), the inclusion of a realistic noise model can create challenges for most optimisation techniques. Thus for our realistic \verb'ibmq_jakarta' noise model calculations, we used the NFT optimiser \citep{NFTopt} instead, as it leads to better convergence. Fig.~\ref{fig:rawenergies-321g-noisy} shows the statistical performance of the HEA[$n_\ell$] ansatz for H$_2$ at $r=1.481696$~{\AA} using the NFT optimiser. Despite the near-converged results of the HEA[3] ansatz in noiseless simulations, we found that the depth of the resulting circuit leads to a very large spread for the measured correlation results (see third set of box plots on the top panel of Fig.~\ref{fig:rawenergies-321g-noisy}). We opted instead for a more compact HEA[1] ansatz as best compromise between circuit depth, correlation description and speed. This is also shown in the lower panel of Fig.~\ref{fig:rawenergies-321g-noisy} where choosing $n_\ell=1$ leads to the least noisy estimation for $E_\mathrm{corr.}$

\subsubsection{Error mitigation for correlation energy}
\begin{figure}[H]
 \centering
 \includegraphics[width=0.9\columnwidth]{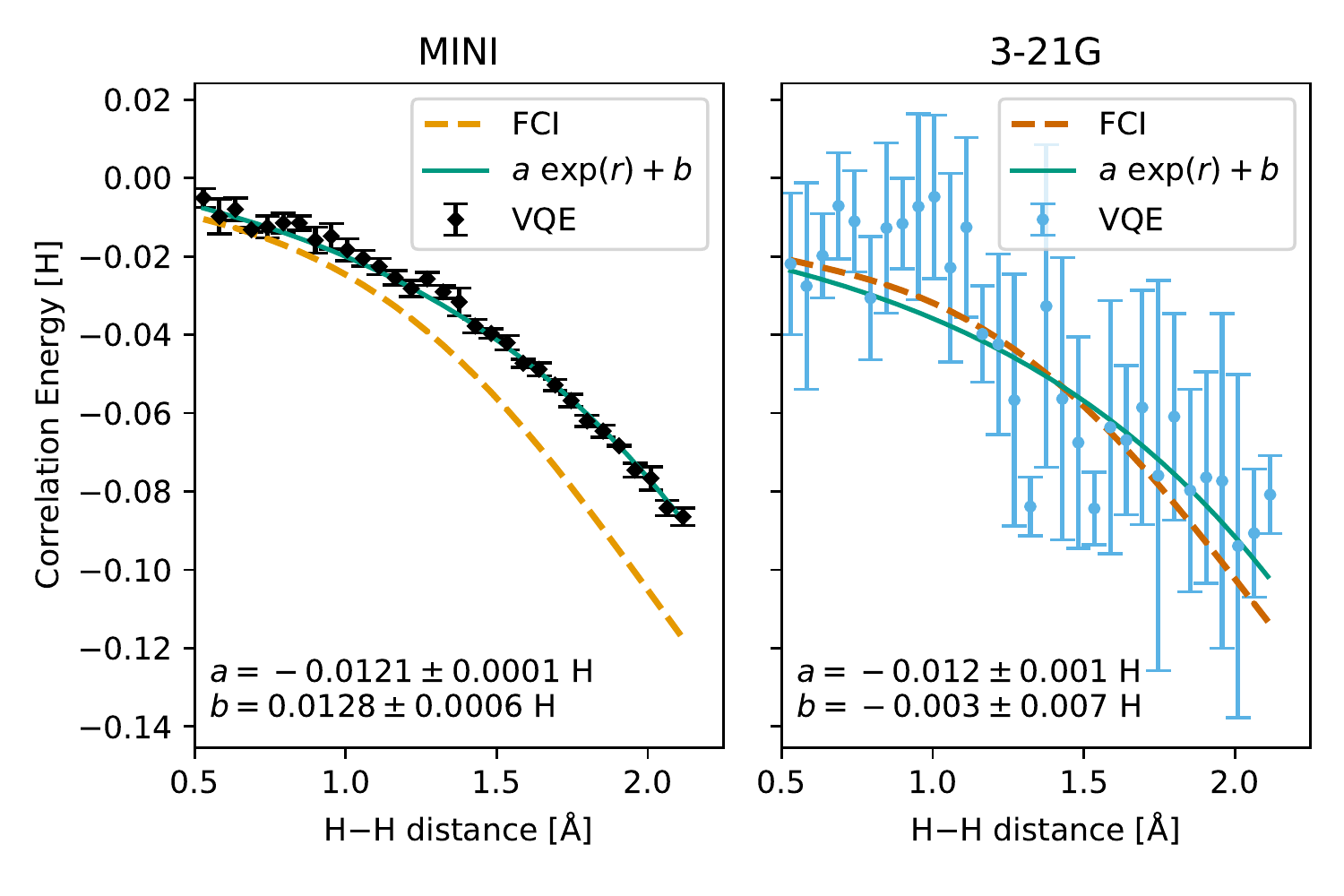}
 \caption{Electronic correlation energy curve for the H$_2$ molecule as a function of H--H separation obtained through VQE simulated experiments using HEA[1]/HFHEA[1] ansatzes for the MINI (diamond) and 3-21G (circles) basis sets. The symbol indicates the median for each series of experiment, computed using the procedure described in Sec.~\ref{sec:corrhfvqe}, along with error bars to indicate 1 standard deviation ($1\sigma$). An empirical fit of the correlation energy to $E_\mathrm{corr.}(r)= a\exp(r)+b$ is also shown as a full green line, along with the fit parameters and their standard deviation at the bottom of each panel. Classically computed curves at the FCI level are also provided for each basis set.}
 \label{fig:noisycorrfig}
\end{figure}

In order to mitigate the remaining error stemming from the determination of the correlation energy using two sets of VQE estimates, we fit the computed correlation energy curves for each basis set to a functional form. Indeed, we saw both in classical and quantum data that the correlation energy displays a monotonous decrease with inter-nuclear separation for this system. Unfortunately, a theoretical functional form that describes the behaviour of the correlation energy is currently not available, so instead we use symbolic regression (PySR package \citep{pysr, cranmer2020discovering}) to build a suitable function. The best scoring function is obtained by training on the correlation data shown in Fig.~\ref{fig:extrapcorr} (MINI, 3-21G and from \citet{sims:2006}). The expression that combines expressiveness with few parameters is determined as $E_\mathrm{corr.}(r)= a\exp(r)+b$. Indeed, this expression fits all classical reference curves with an $R^2\approx 0.99$ and provides a very robust fit for our noisy VQE values. The resulting fits are shown in Fig.~\ref{fig:noisycorrfig} and describe the VQE values well for both MINI basis set ($R^2=0.994$) and 3-21G basis set ($R^2=0.703$) data.

We note however that using a realistic noise model in the simulations caused a marked deviation for the MINI data for large values of $r$ compared to the curve obtained for this basis set in earlier simulations (Fig.~\ref{fig:h2mini321gcorr}). This could be caused by the different ansatz used (HEA[1] vs UCCSD) and the slightly different mapping (BK-parity vs BK). 

\subsection{Full electronic curve}
\begin{figure}[H]
 \centering
 \includegraphics[width=\columnwidth]{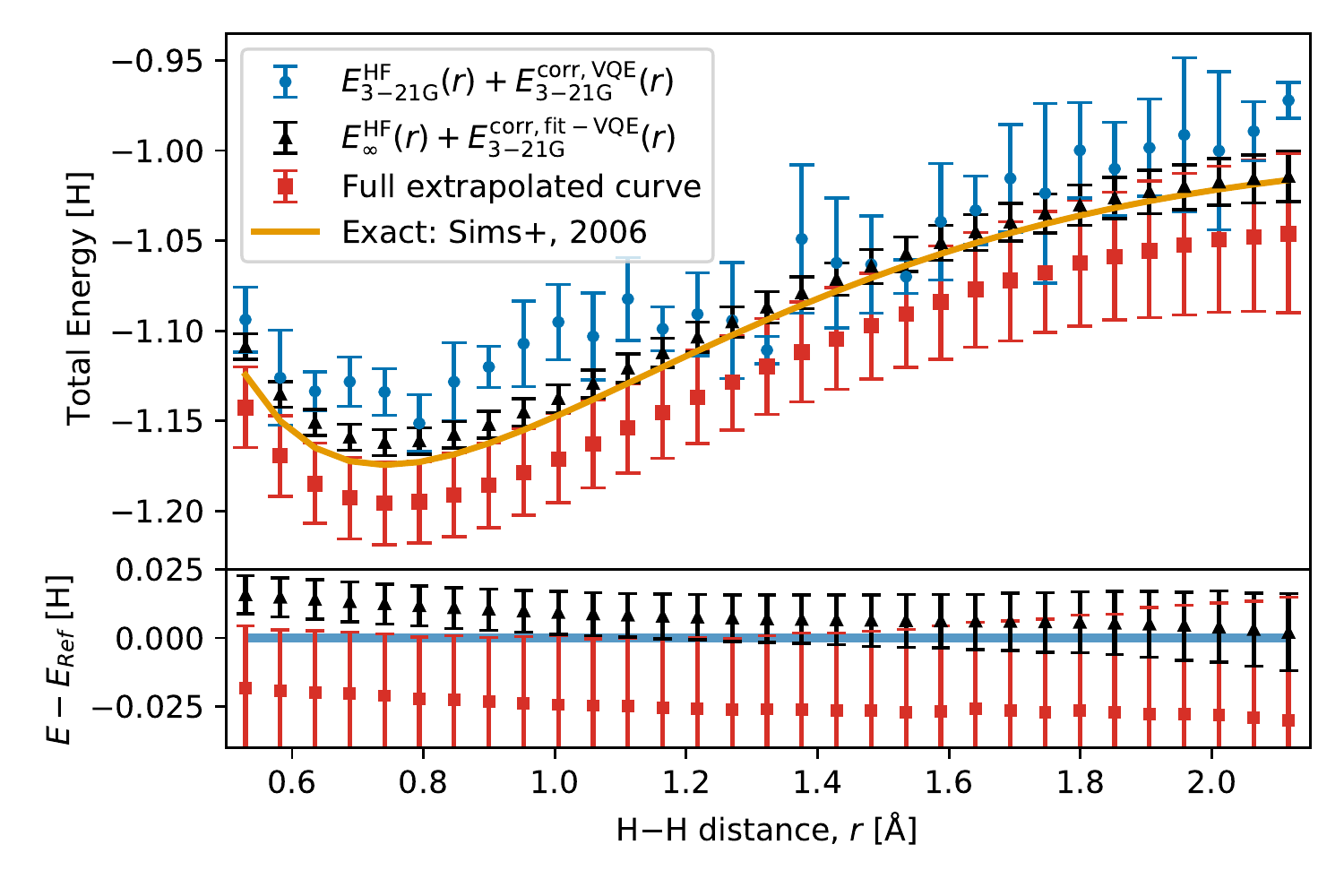}
 \caption{Total electronic energy of H$_2$ as a function of H--H separation, $r$, in Angstroms. The curve is obtained through a mixed (internal standard) approach that uses a HEA[1]/HFHEA[1] combination, along with error mitigation (see text for details). The reconstructed 3-21G VQE estimations are shown as blue circles (Classical $E^\textrm{HF}_\mathrm{3\!-\!21G}$ with VQE $E_\mathrm{corr.}$), an alternative approximation using $E^\textrm{HF}_\infty(r)$ and fitted 3-21G VQE correlation energy is shown as black triangles. The full extrapolated curve, using both $E^\textrm{HF}_\infty(r)$ and fitted $E^\textrm{corr}_\infty$, is shown as red squares. The "exact" classical energy curve from \citet{sims:2006} is shown as an orange continuous line. The symbols indicate the median for set of data points and the error bars indicate a $1\sigma$ standard deviation (computed through bootstrap/Monte-Carlo 16-84 percentile for fitted and extrapolated correlation data, see Sec.~\ref{sec:full_curve} for details). The lower panel shows the deviation from the exact curve ($E-E_{Ref}$) for the two best approximations.}
 \label{fig:h2mini321gnoisy}
\end{figure}

Following the same procedure as for the generic noise simulations (see Sec.~\ref{sec:full_curve}), we now combine the extrapolated HF results from Sec.~\ref{sec:hf_curve} to correlation measurements performed using the "internal standard" approach described earlier (Sec.~\ref{sec:corrhfvqe}). We choose an HEA[1]/HFHEA[1] ansatz combination and a BK-parity mapping as the best compromise to minimise hardware error. As before, we also combine the correlation energy results obtained by fitting the data for both basis sets and use it in the extrapolation equations from Eq.~\ref{eq:correnergy}. The resulting total potential energy curve for H$_2$ obtained for simulations with a realistic noise is shown in Fig.~\ref{fig:h2mini321gnoisy}, along with the reference curve of \citet{sims:2006}.

First of all, to provide a comparison with the generic noise model, we construct a total energy curve for the 3-21G basis set by combining the VQE correlation energy obtained using our error-mitigated HEA[1]/HFHEA[1] approach with the classical Hartree--Fock energy for this basis set, $E^\textrm{HF}_\mathrm{3\!-\!21G}(r)$. As observed previously for our generic noise simulations, 3-21G alone is not able to provide a reliable description of the total energy of the system, although it displays a relatively small VQE error along the curve. More quantitatively, we observe a root mean square deviation (RMDS) from the exact curve of 41~mH over the interaction region (between 0.6~{\AA} and 1.0~{\AA}).

We obtain a much better agreement if instead we use the fitted 3-21G correlation energy obtained for the HEA[1]/HFHEA[1] experiments and combine it with the extrapolated classical HF values, $E^\textrm{HF}_\infty(r)$. We saw previously that this approach gave a very reasonable agreement for the generic noise simulations and we see in Fig.~\ref{fig:h2mini321gnoisy} that this remains the case here too. The generated curve is this time very close to the exact curve and benefits from the smoothing offered by the empirical fitting of the correlation values, which also shrinks the error bars. This approach shows an impressive RMSD from the exact curve of 12~mH for the interaction region. This is likely a fortuitous improvement since the RMSD from the exact curve for the classical FCI/3-21G total energy is 26~mH.

Finally, contrarily to our earlier generic noise simulation results, we see that the extrapolation procedure generates a curve that systematically over-estimate the exact energy curve. This is a likely consequence of the unexpectedly poor description of the MINI correlation energy for this combination of ansatz and mapping in our noisy simulations. While it might be tempting to prefer the earlier approach that combines $E^\textrm{HF}_\infty(r)$ with fitted $E^\textrm{corr,VQE}_\mathrm{3-21G}(r)$, it is unlikely that the correlation obtained with such small basis set would be suitable for larger systems. However, if we consider the lower panel of Fig.~\ref{fig:h2mini321gnoisy}, we observe that both approaches agree with the exact curve within a $1\sigma$ deviation. Lastly, we compute an RMSD from the exact curve in the interaction region is 22~mH for our extrapolated method, which is 9~mH higher than for the generic noise simulations. Yet, given the simplicity of the approach used, it is encouraging that NISQ simulations can provide this level of accuracy. Moreover, any improvements in the VQE approach, such as ADAPT-VQE \citep{Grimsley2019} or iQCC-VQE \citep{Ryabinkin:2020}, to name a few, would automatically improve the accuracy of our technique. 

\section{Conclusions}
\label{sec:conc}
We have shown that our mitigation procedures, combined with an energy extrapolation scheme that uses only sub-minimal basis sets, are able to deliver total energy curves on-par with the best classical data available for the H$_2$ molecule. Indeed, our approach provides data that can theoretically surpass a "chemically-relevant" calculation \citep{Elfving:2020} for this molecule, such as for example a CCSD/cc-pVTZ calculation. We observed that the quality of the final results is largely determined by measurement noise and ansatz quality. In particular, we see that even small basis set calculations can be challenging on simulated NISQ devices. Nevertheless, using an estimation of the reference Hartree--Fock ground state and correlation energy fitting enables a robust mitigation technique that is easily applicable to other VQE approaches or ansatzes.

The extrapolation scheme suggested by \cite{Varandas2018} is very accurate when using classical FCI values (see for example Fig.~\ref{fig:extrapcorr}) and reasonably practical for VQE data, as long as they have a good fidelity (see the MINI basis set data for example). This led to both generic noise simulations and realistic noise simulation data with extrapolation usually improving over the correlation energy computed using the larger 3-21G basis set.

We showed a systematic convergence beyond chemical accuracy for the HEA[$n_\ell$] ansatz in noiseless simulations. However, we observed that a conservative (and less converged) ansatz is preferable in simulations that implement a more realistic noise model. The simple ansatz and a straightforward qubit mapping we use in this work can be advantageously replaced by more elaborate schemes, if a suitable equivalent HF ansatz can be formulated. Further work is underway to explore the application of this scheme to larger systems.

\section*{Acknowledgements}
We thank Zapata Computing Inc.\ for access to the Orquestra software, run-time platform and sponsorship of this work. We acknowledge the Viper High Performance Computing facility of the University of Hull and its support team. We also acknowledge IBMQ education program for providing access to QPU noise models and resources for some early work.

\bibliographystyle{abbrvnat}
\bibliography{M335}

\begin{thebibliography}{49}
\providecommand{\natexlab}[1]{#1}
\providecommand{\url}[1]{\texttt{#1}}
\expandafter\ifx\csname urlstyle\endcsname\relax
  \providecommand{\doi}[1]{doi: #1}\else
  \providecommand{\doi}{doi: \begingroup \urlstyle{rm}\Url}\fi

\bibitem[Bakowies(2007)]{Bakowies:2007}
D.~Bakowies.
\newblock Extrapolation of electron correlation energies to finite and complete
  basis set targets.
\newblock \emph{The Journal of Chemical Physics}, 127\penalty0 (8):\penalty0
  084105, 2007.
\newblock \doi{10.1063/1.2749516}.

\bibitem[{Barlow}(2003)]{Barlow:2003}
R.~{Barlow}.
\newblock {Asymmetric Errors}.
\newblock In L.~{Lyons}, R.~{Mount}, and R.~{Reitmeyer}, editors,
  \emph{Statistical Problems in Particle Physics, Astrophysics, and Cosmology},
  page 250, Jan. 2003.

\bibitem[{Benfenati} et~al.(2021){Benfenati}, {Mazzola}, {Capecci},
  {Barkoutsos}, {Ollitrault}, {Tavernelli}, and {Guidoni}]{benfenati:2021}
F.~{Benfenati}, G.~{Mazzola}, C.~{Capecci}, P.~K. {Barkoutsos}, P.~J.
  {Ollitrault}, I.~{Tavernelli}, and L.~{Guidoni}.
\newblock {Improved accuracy on noisy devices by non-unitary Variational
  Quantum Eigensolver for chemistry applications}.
\newblock \emph{arXiv e-prints}, art. arXiv:2101.09316, Jan. 2021.

\bibitem[Binkley et~al.(1980)Binkley, Pople, and Hehre]{binkley1980a}
J.~S. Binkley, J.~A. Pople, and W.~J. Hehre.
\newblock Self-consistent molecular orbital methods. 21. small split-valence
  basis sets for first-row elements.
\newblock \emph{J. Am. Chem. Soc.}, 102:\penalty0 939--947, 1980.
\newblock \doi{10.1021/ja00523a008}.

\bibitem[Boschen et~al.(2017)Boschen, Theis, Ruedenberg, and
  Windus]{Boschen:2017}
J.~S. Boschen, D.~Theis, K.~Ruedenberg, and T.~L. Windus.
\newblock Correlation energy extrapolation by many-body expansion.
\newblock \emph{The Journal of Physical Chemistry A}, 121\penalty0
  (4):\penalty0 836--844, 2017.
\newblock \doi{10.1021/acs.jpca.6b10953}.

\bibitem[Brandenburg et~al.(2018)Brandenburg, Bannwarth, Hansen, and
  Grimme]{Brandenburg:2018}
J.~G. Brandenburg, C.~Bannwarth, A.~Hansen, and S.~Grimme.
\newblock B97-3c: A revised low-cost variant of the b97-d density functional
  method.
\newblock \emph{The Journal of Chemical Physics}, 148\penalty0 (6):\penalty0
  064104, 2018.
\newblock \doi{10.1063/1.5012601}.

\bibitem[Bravyi et~al.(2017)Bravyi, Gambetta, Mezzacapo, and
  Temme]{bravyi2017tapering}
S.~Bravyi, J.~M. Gambetta, A.~Mezzacapo, and K.~Temme.
\newblock Tapering off qubits to simulate fermionic hamiltonians.
\newblock \emph{arXiv preprint arXiv:1701.08213}, 2017.

\bibitem[Bravyi and Kitaev(2002)]{BRAVYI2002210}
S.~B. Bravyi and A.~Y. Kitaev.
\newblock Fermionic quantum computation.
\newblock \emph{Annals of Physics}, 298\penalty0 (1):\penalty0 210--226, 2002.
\newblock ISSN 0003-4916.
\newblock \doi{10.1006/aphy.2002.6254}.

\bibitem[{Claudino} et~al.(2020){Claudino}, {Wright}, {McCaskey}, and
  {Humble}]{Claudino:2020}
D.~{Claudino}, J.~{Wright}, A.~J. {McCaskey}, and T.~S. {Humble}.
\newblock {Benchmarking Adaptive Variational Quantum Eigensolvers}.
\newblock \emph{Frontiers in Chemistry}, 8:\penalty0 1152, Dec. 2020.
\newblock \doi{10.3389/fchem.2020.606863}.

\bibitem[Cranmer(2020)]{pysr}
M.~Cranmer.
\newblock Pysr: Fast \& parallelized symbolic regression in python/julia, Sept.
  2020.
\newblock URL \url{http://doi.org/10.5281/zenodo.4041459}.

\bibitem[Cranmer et~al.(2020)Cranmer, Sanchez-Gonzalez, Battaglia, Xu, Cranmer,
  Spergel, and Ho]{cranmer2020discovering}
M.~Cranmer, A.~Sanchez-Gonzalez, P.~Battaglia, R.~Xu, K.~Cranmer, D.~Spergel,
  and S.~Ho.
\newblock Discovering symbolic models from deep learning with inductive biases.
\newblock \emph{NeurIPS 2020}, 2020.

\bibitem[Dean and Dixon(1951)]{Dean:1951}
R.~B. Dean and W.~J. Dixon.
\newblock Simplified statistics for small numbers of observations.
\newblock \emph{Analytical Chemistry}, 23\penalty0 (4):\penalty0 636--638,
  1951.
\newblock \doi{10.1021/ac60052a025}.

\bibitem[Dunning(1989)]{dunning:1989}
T.~H. Dunning.
\newblock Gaussian basis sets for use in correlated molecular calculations. i.
  the atoms boron through neon and hydrogen.
\newblock \emph{The Journal of Chemical Physics}, 90\penalty0 (2):\penalty0
  1007--1023, 1989.
\newblock \doi{10.1063/1.456153}.

\bibitem[Echenique and Alonso(2007)]{Echenique:2007}
P.~Echenique and J.~L. Alonso.
\newblock A mathematical and computational review of hartree–fock scf methods
  in quantum chemistry.
\newblock \emph{Molecular Physics}, 105\penalty0 (23-24):\penalty0 3057--3098,
  2007.
\newblock \doi{10.1080/00268970701757875}.

\bibitem[{Elfving} et~al.(2020){Elfving}, {Broer}, {Webber}, {Gavartin},
  {Halls}, {Lorton}, and {Bochevarov}]{Elfving:2020}
V.~E. {Elfving}, B.~W. {Broer}, M.~{Webber}, J.~{Gavartin}, M.~D. {Halls},
  K.~P. {Lorton}, and A.~{Bochevarov}.
\newblock {How will quantum computers provide an industrially relevant
  computational advantage in quantum chemistry?}
\newblock \emph{arXiv e-prints}, art. arXiv:2009.12472, Sept. 2020.

\bibitem[Feller(1996)]{feller1996a}
D.~Feller.
\newblock The role of databases in support of computational chemistry
  calculations.
\newblock \emph{J. Comput. Chem.}, 17:\penalty0 1571--1586, 1996.
\newblock \doi{10.1002/(SICI)1096-987X(199610)17:13<1571::AID-JCC9>3.0.CO;2-P}.

\bibitem[{Google~Inc.}(2022)]{openfermion}
{Google~Inc.}
\newblock Openfermion, 2022.
\newblock URL \url{https://quantumai.google/openfermion}.

\bibitem[Grimme et~al.(2015)Grimme, Brandenburg, Bannwarth, and
  Hansen]{grimme:2015}
S.~Grimme, J.~G. Brandenburg, C.~Bannwarth, and A.~Hansen.
\newblock Consistent structures and interactions by density functional theory
  with small atomic orbital basis sets.
\newblock \emph{The Journal of Chemical Physics}, 143\penalty0 (5):\penalty0
  054107, 2015.
\newblock \doi{10.1063/1.4927476}.

\bibitem[Grimsley et~al.(2019)Grimsley, Economou, Barnes, and
  Mayhall]{Grimsley2019}
H.~R. Grimsley, S.~E. Economou, E.~Barnes, and N.~J. Mayhall.
\newblock An adaptive variational algorithm for exact molecular simulations on
  a quantum computer.
\newblock \emph{Nature Communications}, 10\penalty0 (1):\penalty0 3007, Jul
  2019.
\newblock ISSN 2041-1723.
\newblock \doi{10.1038/s41467-019-10988-2}.

\bibitem[Halkier et~al.(1998)Halkier, Helgaker, J{\o}rgensen, Klopper, Koch,
  Olsen, and Wilson]{Halkier:1998}
A.~Halkier, T.~Helgaker, P.~J{\o}rgensen, W.~Klopper, H.~Koch, J.~Olsen, and
  A.~K. Wilson.
\newblock Basis-set convergence in correlated calculations on ne, n2, and h2o.
\newblock \emph{Chemical Physics Letters}, 286\penalty0 (3):\penalty0 243--252,
  1998.
\newblock ISSN 0009-2614.
\newblock \doi{10.1016/S0009-2614(98)00111-0}.

\bibitem[Hedges and Shah(2003)]{hedges:2003}
S.~Hedges and P.~Shah.
\newblock Comparison of mode estimation methods and application in molecular
  clock analysis.
\newblock \emph{BMC Bioinformatics}, 4:\penalty0 31, 2003.
\newblock \doi{10.1186/1471-2105-4-31}.

\bibitem[Hughes and Hase(2010)]{errorbook}
I.~Hughes and T.~Hase.
\newblock \emph{Measurements and their Uncertainties}.
\newblock Oxford University Press, UK, 2010.
\newblock ISBN 9780199566334.

\bibitem[Jensen(2001)]{Jensen:2001}
F.~Jensen.
\newblock Polarization consistent basis sets: Principles.
\newblock \emph{The Journal of Chemical Physics}, 115\penalty0 (20):\penalty0
  9113--9125, 2001.
\newblock \doi{10.1063/1.1413524}.

\bibitem[Jensen(2005)]{Jensen2005}
F.~Jensen.
\newblock Estimating the hartree---fock limit from finite basis set
  calculations.
\newblock \emph{Theoretical Chemistry Accounts}, 113\penalty0 (5):\penalty0
  267--273, Jun 2005.
\newblock ISSN 1432-2234.
\newblock \doi{10.1007/s00214-005-0635-2}.

\bibitem[{Jouzdani} and {Bringuier}(2020)]{jouzdani:2020}
P.~{Jouzdani} and S.~{Bringuier}.
\newblock {Hybrid Quantum-Classical Eigensolver Without Variation or Parametric
  Gates}.
\newblock \emph{arXiv e-prints}, art. arXiv:2008.11347, Aug. 2020.

\bibitem[Kandala et~al.(2017)Kandala, Mezzacapo, Temme, Takita, Brink, Chow,
  and Gambetta]{Kandala2017}
A.~Kandala, A.~Mezzacapo, K.~Temme, M.~Takita, M.~Brink, J.~M. Chow, and J.~M.
  Gambetta.
\newblock Hardware-efficient variational quantum eigensolver for small
  molecules and quantum magnets.
\newblock \emph{Nature}, 549\penalty0 (7671):\penalty0 242--246, Sep 2017.
\newblock ISSN 1476-4687.
\newblock \doi{10.1038/nature23879}.

\bibitem[{Kottmann} et~al.(2020){Kottmann}, {Schleich}, {Tamayo-Mendoza}, and
  {Aspuru-Guzik}]{kottmann:2020}
J.~S. {Kottmann}, P.~{Schleich}, T.~{Tamayo-Mendoza}, and A.~{Aspuru-Guzik}.
\newblock {Reducing qubit requirements while maintaining numerical precision
  for the Variational Quantum Eigensolver: A Basis-Set-Free Approach}.
\newblock \emph{arXiv e-prints}, art. arXiv:2008.02819, Aug. 2020.

\bibitem[{Laursen} et~al.(2019){Laursen}, {Sommer-Larsen}, {Milvang-Jensen},
  {Fynbo}, and {Razoumov}]{laursen_error:2019}
P.~{Laursen}, J.~{Sommer-Larsen}, B.~{Milvang-Jensen}, J.~P.~U. {Fynbo}, and
  A.~O. {Razoumov}.
\newblock {Lyman {\ensuremath{\alpha}}-emitting galaxies in the epoch of
  reionization}.
\newblock \emph{\aap}, 627:\penalty0 A84, July 2019.
\newblock \doi{10.1051/0004-6361/201833645}.

\bibitem[{Mitin}(2000)]{mitin2000}
A.~V. {Mitin}.
\newblock {Exact solution of the Hartree-Fock equation for the H$_{2}$ molecule
  in the linear-combination-of-atomic-orbitals approximation}.
\newblock \emph{\pra}, 62\penalty0 (1):\penalty0 010501, July 2000.
\newblock \doi{10.1103/PhysRevA.62.010501}.

\bibitem[Nakanishi et~al.(2020)Nakanishi, Fujii, and Todo]{NFTopt}
K.~M. Nakanishi, K.~Fujii, and S.~Todo.
\newblock Sequential minimal optimization for quantum-classical hybrid
  algorithms.
\newblock \emph{Phys. Rev. Research}, 2:\penalty0 043158, Oct 2020.
\newblock \doi{10.1103/PhysRevResearch.2.043158}.

\bibitem[Neese(2012)]{neese_2012}
F.~Neese.
\newblock The orca program system.
\newblock \emph{WIREs Comput. Mol. Sci.}, 2\penalty0 (1):\penalty0 73--78,
  2012.
\newblock \doi{10.1002/wcms.81}.

\bibitem[{Pachucki}(2010)]{Pachucki:2010}
K.~{Pachucki}.
\newblock {Born-Oppenheimer potential for H$_{2}$}.
\newblock \emph{\pra}, 82\penalty0 (3):\penalty0 032509, Sept. 2010.
\newblock \doi{10.1103/PhysRevA.82.032509}.

\bibitem[{Peruzzo} et~al.(2014){Peruzzo}, {McClean}, {Shadbolt}, {Yung},
  {Zhou}, {Love}, {Aspuru-Guzik}, and {O'Brien}]{Peruzzo:2014}
A.~{Peruzzo}, J.~{McClean}, P.~{Shadbolt}, M.-H. {Yung}, X.-Q. {Zhou}, P.~J.
  {Love}, A.~{Aspuru-Guzik}, and J.~L. {O'Brien}.
\newblock {A variational eigenvalue solver on a photonic quantum processor}.
\newblock \emph{Nature Communications}, 5:\penalty0 4213, July 2014.
\newblock \doi{10.1038/ncomms5213}.

\bibitem[Powell(1964)]{powell1964}
M.~J.~D. Powell.
\newblock {An efficient method for finding the minimum of a function of several
  variables without calculating derivatives}.
\newblock \emph{The Computer Journal}, 7\penalty0 (2):\penalty0 155--162, 01
  1964.
\newblock ISSN 0010-4620.
\newblock \doi{10.1093/comjnl/7.2.155}.

\bibitem[Press et~al.(2007)Press, Teukolsky, Vetterling, and
  Flannery]{numericalrecipes}
W.~H. Press, S.~A. Teukolsky, W.~T. Vetterling, and B.~P. Flannery.
\newblock \emph{Numerical Recipes 3rd Edition: The Art of Scientific
  Computing}.
\newblock Cambridge University Press, USA, 3 edition, 2007.
\newblock ISBN 0521880688.

\bibitem[Pritchard et~al.(2019)Pritchard, Altarawy, Didier, Gibsom, and
  Windus]{pritchard2019a}
B.~P. Pritchard, D.~Altarawy, B.~Didier, T.~D. Gibsom, and T.~L. Windus.
\newblock A new basis set exchange: An open, up-to-date resource for the
  molecular sciences community.
\newblock \emph{J. Chem. Inf. Model.}, 59:\penalty0 4814--4820, 2019.
\newblock \doi{10.1021/acs.jcim.9b00725}.

\bibitem[Romano(1988)]{romano:88}
J.~P. Romano.
\newblock Bootstrapping the mode.
\newblock \emph{Ann. Inst. Statist. Math.}, 40:\penalty0 565--586, 1988.

\bibitem[Ryabinkin et~al.(2020)Ryabinkin, Lang, Genin, and
  Izmaylov]{Ryabinkin:2020}
I.~G. Ryabinkin, R.~A. Lang, S.~N. Genin, and A.~F. Izmaylov.
\newblock Iterative qubit coupled cluster approach with efficient screening of
  generators.
\newblock \emph{Journal of Chemical Theory and Computation}, 16\penalty0
  (2):\penalty0 1055--1063, 2020.
\newblock \doi{10.1021/acs.jctc.9b01084}.

\bibitem[{Sajid~Anis} et~al.(2021){Sajid~Anis}, Abby-Mitchell, and
  Abraham]{shortQiskit}
M.~{Sajid~Anis}, Abby-Mitchell, and H.~Abraham.
\newblock Qiskit: An open-source framework for quantum computing, 2021.

\bibitem[Schuchardt et~al.(2007)Schuchardt, Didier, Elsethagen, Sun,
  Gurumoorthi, Chase, Li, and Windus]{schuchardt2007a}
K.~L. Schuchardt, B.~T. Didier, T.~Elsethagen, L.~Sun, V.~Gurumoorthi,
  J.~Chase, J.~Li, and T.~L. Windus.
\newblock Basis set exchange: A community database for computational sciences.
\newblock \emph{J. Chem. Inf. Model.}, 47:\penalty0 1045--1052, 2007.
\newblock \doi{10.1021/ci600510j}.

\bibitem[Seeley et~al.(2012)Seeley, Richard, and Love]{Seeley2012}
J.~T. Seeley, M.~J. Richard, and P.~J. Love.
\newblock The bravyi-kitaev transformation for quantum computation of
  electronic structure.
\newblock \emph{The Journal of Chemical Physics}, 137\penalty0 (22):\penalty0
  224109, 2012.
\newblock \doi{10.1063/1.4768229}.

\bibitem[{Sims} and {Hagstrom}(2006)]{sims:2006}
J.~S. {Sims} and S.~A. {Hagstrom}.
\newblock {High precision variational calculations for the Born-Oppenheimer
  energies of the ground state of the hydrogen molecule}.
\newblock \emph{\jcp}, 124\penalty0 (9):\penalty0 094101--094101, Mar. 2006.
\newblock \doi{10.1063/1.2173250}.

\bibitem[Sure and Grimme(2013)]{Sure:2013}
R.~Sure and S.~Grimme.
\newblock Corrected small basis set hartree-fock method for large systems.
\newblock \emph{Journal of Computational Chemistry}, 34\penalty0 (19):\penalty0
  1672--1685, 2013.
\newblock \doi{10.1002/jcc.23317}.

\bibitem[Taube and Bartlett(2006)]{Taube2006}
A.~G. Taube and R.~J. Bartlett.
\newblock New perspectives on unitary coupled-cluster theory.
\newblock \emph{International Journal of Quantum Chemistry}, 106\penalty0
  (15):\penalty0 3393--3401, 2006.
\newblock \doi{10.1002/qua.21198}.

\bibitem[Turney et~al.(2012)Turney, Simmonett, Parrish, Hohenstein,
  Evangelista, Fermann, Mintz, Burns, Wilke, Abrams, Russ, Leininger, Janssen,
  Seidl, Allen, Schaefer, King, Valeev, Sherrill, and Crawford]{psi4:2012}
J.~M. Turney, A.~C. Simmonett, R.~M. Parrish, E.~G. Hohenstein, F.~A.
  Evangelista, J.~T. Fermann, B.~J. Mintz, L.~A. Burns, J.~J. Wilke, M.~L.
  Abrams, N.~J. Russ, M.~L. Leininger, C.~L. Janssen, E.~T. Seidl, W.~D. Allen,
  H.~F. Schaefer, R.~A. King, E.~F. Valeev, C.~D. Sherrill, and T.~D. Crawford.
\newblock Psi4: an open-source ab initio electronic structure program.
\newblock \emph{WIREs Computational Molecular Science}, 2\penalty0
  (4):\penalty0 556--565, 2012.
\newblock \doi{10.1002/wcms.93}.

\bibitem[van Duijneveldt(1971)]{duijneveldt1971a}
F.~B. van Duijneveldt.
\newblock Gaussian basis sets for the atoms h-ne for use in molecular
  calculations.
\newblock Technical Report RJ 945, IBM Research Laboratory, San Jose,
  California, 1971.

\bibitem[Varandas(2018)]{Varandas2018}
A.~J.~C. Varandas.
\newblock {CBS} extrapolation in electronic structure pushed to the end: a
  revival of minimal and sub-minimal basis sets.
\newblock \emph{Phys. Chem. Chem. Phys.}, 20:\penalty0 22084--22098, 2018.
\newblock \doi{10.1039/C8CP02932F}.

\bibitem[Virtanen et~al.(2020)Virtanen, Gommers, Oliphant, Haberland, Reddy,
  Cournapeau, Burovski, Peterson, Weckesser, Bright, {van der Walt}, Brett,
  Wilson, Millman, Mayorov, Nelson, Jones, Kern, Larson, Carey, Polat, Feng,
  Moore, {VanderPlas}, Laxalde, Perktold, Cimrman, Henriksen, Quintero, Harris,
  Archibald, Ribeiro, Pedregosa, {van Mulbregt}, and {SciPy 1.0
  Contributors}]{2020SciPy-NMeth}
P.~Virtanen, R.~Gommers, T.~E. Oliphant, M.~Haberland, T.~Reddy, D.~Cournapeau,
  E.~Burovski, P.~Peterson, W.~Weckesser, J.~Bright, S.~J. {van der Walt},
  M.~Brett, J.~Wilson, K.~J. Millman, N.~Mayorov, A.~R.~J. Nelson, E.~Jones,
  R.~Kern, E.~Larson, C.~J. Carey, {\.I}.~Polat, Y.~Feng, E.~W. Moore,
  J.~{VanderPlas}, D.~Laxalde, J.~Perktold, R.~Cimrman, I.~Henriksen, E.~A.
  Quintero, C.~R. Harris, A.~M. Archibald, A.~H. Ribeiro, F.~Pedregosa, P.~{van
  Mulbregt}, and {SciPy 1.0 Contributors}.
\newblock {{SciPy} 1.0: Fundamental Algorithms for Scientific Computing in
  Python}.
\newblock \emph{Nature Methods}, 17:\penalty0 261--272, 2020.
\newblock \doi{10.1038/s41592-019-0686-2}.

\bibitem[{Zapata~Inc.}(2021)]{orquestra}
{Zapata~Inc.}
\newblock Orquestra, 2021.
\newblock URL \url{http://www.zapatacomputing.com/orquestra}.

\end{thebibliography}

\end{document}